\documentclass[preprint,authoryear,12pt]{elsarticle}
\usepackage{graphics}
\usepackage{graphicx}
\usepackage{epsfig}

\usepackage{amssymb}
\usepackage{amsthm}

\usepackage{lineno}

\journal{International Journal of Engineering Science}

\begin{document}

\begin{frontmatter}



\title{
%Anisotropy and Anomalous Scaling in 
Steady Homogeneous Turbulence in the Presence of an  Average Velocity Gradient}


\author{Nicola de Divitiis}

\address{"La Sapienza" University, Dipartimento di Ingegneria Meccanica e Aerospaziale, Via Eudossiana, 18, 00184 Rome, Italy}

\begin{abstract}
We study the homogeneous turbulence in the presence of a constant average velocity
gradient in an infinite fluid domain, with a novel finite-scale Lyapunov analysis, 
presented in a previous work dealing with the homogeneous isotropic turbulence.

Here, the energy spectrum is studied introducing the spherical averaged pair correlation function, whereas the anisotropy caused by the velocity gradient is analyzed using the equation
of the two points velocity distribution function which is determined through the Liouville theorem.
As a result, we obtain the evolution equation of this velocity correlation function
which is shown to be valid also when the fluid motion is referred with respect to a rotating
reference frame. This equation tends to the classical von K\'arm\'an-Howarth equation when the average velocity gradient vanishes.

We show that, the steady energy spectrum, instead of following the Kolmogorov law $\kappa^{-5/3}$, varies as $\kappa^{-2}$. Accordingly, the structure function of the longitudinal velocity difference $\langle \Delta u_r^n \rangle \approx r^{\zeta_n}$ exhibits the anomalous scaling $\zeta_n \approx n/2$, and the integral scales of the correlation function are much smaller than those of the isotropic turbulence.
\end{abstract}


\end{frontmatter}

\newcommand{\no}{\noindent}
\newcommand{\be}{\begin{equation}}
\newcommand{\ee}{\end{equation}}
\newcommand{\bea}{\begin{eqnarray}}
\newcommand{\eea}{\end{eqnarray}}
\newcommand{\bc}{\begin{center}}
\newcommand{\ec}{\end{center}}

\newcommand{\calr}{{\cal R}}
\newcommand{\calv}{{\cal V}}

\newcommand{\bff}{\mbox{\boldmath $f$}}
\newcommand{\bfg}{\mbox{\boldmath $g$}}
\newcommand{\bfh}{\mbox{\boldmath $h$}}
\newcommand{\bfi}{\mbox{\boldmath $i$}}
\newcommand{\bfm}{\mbox{\boldmath $m$}}
\newcommand{\bfp}{\mbox{\boldmath $p$}}
\newcommand{\bfr}{\mbox{\boldmath $r$}}
\newcommand{\bfu}{\mbox{\boldmath $u$}}
\newcommand{\bfv}{\mbox{\boldmath $v$}}
\newcommand{\bfx}{\mbox{\boldmath $x$}}
\newcommand{\bfy}{\mbox{\boldmath $y$}}
\newcommand{\bfw}{\mbox{\boldmath $w$}}
\newcommand{\bfk}{\mbox{\boldmath $\kappa$}}

\newcommand{\bfA}{\mbox{\boldmath $A$}}
\newcommand{\bfD}{\mbox{\boldmath $D$}}
\newcommand{\bfI}{\mbox{\boldmath $I$}}
\newcommand{\bfL}{\mbox{\boldmath $L$}}
\newcommand{\bfM}{\mbox{\boldmath $M$}}
\newcommand{\bfS}{\mbox{\boldmath $S$}}
\newcommand{\bfT}{\mbox{\boldmath $T$}}
\newcommand{\bfU}{\mbox{\boldmath $U$}}
\newcommand{\bfX}{\mbox{\boldmath $X$}}
\newcommand{\bfY}{\mbox{\boldmath $Y$}}
\newcommand{\bfK}{\mbox{\boldmath $K$}}

\newcommand{\bfrho}{\mbox{\boldmath $\rho$}}
\newcommand{\bfchi}{\mbox{\boldmath $\chi$}}
\newcommand{\bfphi}{\mbox{\boldmath $\phi$}}
\newcommand{\bfPhi}{\mbox{\boldmath $\Phi$}}
\newcommand{\bflambda}{\mbox{\boldmath $\lambda$}}
\newcommand{\bfxi}{\mbox{\boldmath $\xi$}}
\newcommand{\bfLambda}{\mbox{\boldmath $\Lambda$}}
\newcommand{\bfPsi}{\mbox{\boldmath $\Psi$}}
\newcommand{\bfomega}{\mbox{\boldmath $\omega$}}
\newcommand{\bfOmega}{\mbox{\boldmath $\Omega$}}
\newcommand{\bfeps}{\mbox{\boldmath $\varepsilon$}}
\newcommand{\bfepsn}{\mbox{\boldmath $\epsilon$}}
\newcommand{\bfzeta}{\mbox{\boldmath $\zeta$}}
\newcommand{\bfkappa}{\mbox{\boldmath $\kappa$}}
\newcommand{\itPsi}{\mbox{\it $\Psi$}}
\newcommand{\itPhi}{\mbox{\it $\Phi$}}
\newcommand{\bint}{\mbox{ \int{a}{b}} }
\newcommand{\ds}{\displaystyle}
\newcommand{\Sum}{\Large \sum}

% \linenumbers

\section{Introduction \label{s1}}

%\label{intro}

Although the Kolmogorov law $E(\kappa) \approx k^{-5/3}$ represents the main result
of the isotropic turbulence, there are many experimental evidences and theoretical
arguments indicating that this is not the only spectrum observed in the fully developed turbulence of incompressible fluids (\cite{Frisch73}, \cite{Moffat78}, \cite{Gordienko01}, \cite{Swinney2010}).

For example, in \cite{Frisch73} and \cite{Moffat78}, it is shown through the 
dimensional analysis, that the energy spectrum in the presence of an average velocity gradient $\partial U/\partial y$ can follow the law $\approx k^{-7/3}$ in the inertial subrange.
This is a particular result arising from the assumption that the energy spectrum is linear in $\partial U/\partial y$. More in general, assuming that the energy spectrum is proportional to $(\partial U/\partial y)^\beta$ with  $\beta>0$, the Buckingham theorem states that $E(\kappa) \approx \kappa^{-5/3 -2/3 \beta}$, and different scaling exponent are possible.

\cite{Gordienko01} studied the forced driving turbulence,
where the forcing term can have various origins.
The authors remarked that there are two dimensionless parameters, characterizing the forcing term, which influence the shape of the energy spectrum and are responsible for the anomalous spectra.
They showed that, in a certain interval of variation of one of these parameters, the spectrum follows the Kolmogorov law, whereas for an opportune choice of it, the spectrum behaves like $\kappa^{-2}$ in the inertial subrange.

These different scaling are caused by the shear rate which leads to the development of coherent fluid structures.
These are streaky structures, due to the stretching of the vortex lines, which exhibit the maximum dimension along the stream direction (\cite{Kim}).
In Ref. (\cite{Kim}), the authors remark that these streaky structures influence the spanwise correlation of streamwise velocity components, being a non monotonic function
which becomes negative at high values of the spanwise distances.

Further, in \cite{Swinney2010}, the authors experimentally analyzed the statistics
of the longitudinal velocity difference $\Delta u_r$ in a closed cylindric tank which rotates around its symmetry axis at a given spin rate. The turbulence is generated by pumping the flow in the tank through two concentric rings of 120 holes each, placed at the bottom of the tank, where the source ring is the internal one.
This generates an average radial flow that, combined with the spin rate, determines
a Coriolis force whose magnitude varies with the distance from the rotation axis.
As a consequence, a counterrotating flow and an average velocity gradient with
respect to the tank frame is observed (\cite{Swinney2010}).
The authors found that $\Delta u_r$ presents the anomalous scaling $\langle \Delta u_r^n \rangle \approx r^{\zeta_n}$ with
$\zeta_n \simeq n/2$, in contrast with the Kolmogorov law ($\zeta_n \simeq n/3$),
and that  $E(\kappa) \approx \kappa^{-2}$.

\bigskip

The present work studies the homogeneous turbulence in an infinite fluid domain 
in the presence of an average velocity gradient $\nabla_{\bf x} {\bf U}$, using the finite-scale Lyapunov analysis, proposed by \cite{deDivitiis_1}, \cite{deDivitiis_2}
for studying the homogeneous isotropic turbulence.

In the first section, we define the spherical part of the velocity correlation tensor $R_{i j}$,  and we derive the evolution equation for $R_{i i}$ from the Navier-Stokes equation with $\nabla_{\bf x} {\bf U} \ne \bf 0$.

To study the effect of $\nabla_{\bf x} {\bf U}$ on the anisotropy 
and on $R_{i j}$, the evolution equation for the pair
distribution function is derived from the Liouville theorem, assuming that the
statistical equilibrium corresponds to the condition of isotropic turbulence
when the kinetic energy rate is equal to zero.
From this equation, the steady velocity correlation tensor is expressed in function of the average velocity gradient and of the maximal finite-scale Lyapunov exponent and, in particular, the Boussinesq closure for the Reynolds stress is obtained.

Finally, the equation for the spherical averaged longitudinal correlation function is determined, 
whose solutions depend on the average velocity gradient. 
Moreover, we show that this equation is still valid when the fluid motion is refereed 
with respect to a non-inertial rotating frame of reference.
The steady solutions of this equation are numerically calculated for
different Taylor-Scale Reynolds number and several results are presented.
We found that $E(\kappa) \approx \kappa^{-2}$ in the inertial subrange, thus 
the statistical moments $\langle \Delta u_r^n \rangle \approx r^{\zeta_n}$
exhibit the anomalous scaling $\zeta_n \approx n/2$, whereas the integral scales
of the longitudinal correlation function are much lesser than those of the
isotropic turbulence.
In the case of homogeneous turbulence in the presence of a steady shear rate, 
the spanwise correlation function of the streamwise velocity component is also 
calculated.
% This results to be a non-monotonic function which becomes negative
%for high values of the spanwise separation distance.

\bigskip

\section{\bf  Analysis}

This section analyzes the homogeneous turbulence with an uniform average velocity gradient $\nabla_{\bf x} {\bf U}$.

The fluid velocity, measured in the reference frame $\Re$, is
${\bf v} ={\bf U} + {\bf u}$, where ${\bf U} \equiv (U_x, U_y, U_z)$ and 
${\bf u} \equiv (u_x, u_y, u_z)$ are, average and fluctuating velocity, respectively.
The velocity correlation tensor is defined as $R_{i j} = \langle u_i u'_j \rangle$, being $u_i$ and $u'_j$ the velocity components of $\bf u$ calculated at $\bf x$ 
and ${\bf x'} = {\bf x} + {\bf r}$, where the brackets denote the average on the statistical ensemble of $\bf u$ and $\bf u'$, and $\bf r$ is the separation distance (\cite{Karman38}, \cite{Batchelor53}).

\bigskip

In order to determine the evolution equation of $R_{i j}$, we start from the Navier-Stokes equations, written for the fluctuating velocity (\cite{Batchelor53}), in the points 
${\bf x}$ and ${\bf x}'$
\begin{eqnarray}
\begin{array}{l@{\hspace{+0.2cm}}l}
\ds \frac{\partial u_i}{\partial t}=
-\frac{\partial u_i u_k}{\partial x_k} 
-\frac{\partial U_i u_k}{\partial x_k}
-\frac{\partial u_i U_k}{\partial x_k}
-\frac{1}{\rho}  \frac{\partial p}{\partial x_i}
+ \nu \nabla^2 u_i \\\\
\ds \frac{\partial u'_j}{\partial t}=
-\frac{\partial u'_j u'_k}{\partial x'_k} 
-\frac{\partial U'_j u'_k}{\partial x'_k}
-\frac{\partial u'_j U'_k}{\partial x'_k}
-\frac{1}{\rho}  \frac{\partial p'}{\partial x'_j}
+ \nu \nabla'^2 u'_j
\end{array}
\label{N-S}
\end{eqnarray}
where $p$ is the fluctuating pressure and ${\bf U}'$ is
\bea
{\bf U}' 
=  {\bf U} + \nabla_{\bf x} {\bf U} \ {\bf r}
\label{grad U}
\eea
being ${\bf U}$ and $\nabla_{\bf x} {\bf U}$ assigned quantities.
The repeated index indicates the summation with respect the same index.
The evolution equation of $R_{i j}$ is determined 
by multiplying first and second equation by $u'_j$ and $u_i$, respectively, summing the so obtained equations, and calculating the average on the statistical ensemble (\cite{Batchelor53}):
\bea
\ds
\frac{\partial R_{i j}}{\partial t}=
T_{i j} + P_{i j} 
+ 2 \nu \nabla^2 R_{i j} 
- \frac{\partial U_i}{\partial x_k} R_{k j}
- \frac{\partial U_j}{\partial x_k} R_{i k}  
+ \frac{\partial R_{i j} }{\partial r_k} 
(U_k - U_k')
\label{cc1 A}
\label{cc1}
\eea
being 
\bea
\begin{array}{l@{\hspace{+0.2cm}}l}
\ds T_{i j} ({\bf r}) =\frac{\partial }{\partial r_k} 
\left\langle u_i  u_j' (u_k - u_k')  \right\rangle, \ \ \ \
\ds P_{i j} ({\bf r}) = \frac{1}{\rho} 
\left( \frac{\partial \langle p u'_j \rangle}{\partial r_i} 
- \frac{\partial \langle p' u_i \rangle }{\partial r_j} \right)  
\end{array}
\eea
where $\partial \langle...\rangle /\partial x_i \equiv -\partial\langle...\rangle / \partial r_i$ 
and $\partial \langle...\rangle / \partial x'_i \equiv \partial \langle...\rangle / \partial r_i$. Making the trace of Eq. (\ref{cc1 A}), we obtain the following scalar equation
\bea
\ds
\frac{\partial R}{\partial t}=
\frac{1}{2} H 
+ 2 \nu \nabla^2 R 
- \frac{\partial U_i}{\partial x_k} R_{i k}^S
+ \frac{\partial R }{\partial r_k} 
(U_k - U_k')
\label{cc2}
\eea
where $R_{i k}^S$ is the symmetric part of $R_{i k}$, and  $R$, defined as
\bea
\ds R = \frac{1}{2} {R_{i i}}, 
\eea
 gives the turbulent kinetic energy for ${\bf r}$=$0$, 
$H$ $\equiv$ $T_{i i}$ provides the mechanism of energy cascade, and
$P_{i i} ({\bf r}) \equiv 0$ arises from the fluid incompressibility (\cite{Karman38},  \cite{Batchelor53}).

Equation (\ref{cc2}) incudes two additional terms with respect to the homogeneous isotropic turbulence.  
$- {\partial U_i}/{\partial r_k} R_{i k}$ corresponds to the kinetic energy production if 
$- {\partial U_i}/{\partial r_k} R_{i k}(0) >0$, whereas 
$ {\partial R_{i i} }/{\partial r_k} (U_k - U_k')$ represents an energy transfer as this latter vanishes for ${\bf r} =0$.

\bigskip

Assuming the condition of isotropic turbulence, 
$\partial U_i/\partial x_k R_{k i}$ 
does not give energy production, and the flow corresponds to a dying turbulence whose
dissipation rate is calculated with the von K\'arm\'an-Howarth equation.
In this case the energy spectrum is a function of $\bfkappa^2 \equiv \kappa_i \kappa_i$.

On the contrary, the combined effect of the anisotropy (i.e. 
$\langle u_x^2 \rangle$ $\ne$ $\langle u_y^2 \rangle$,  
$\langle u_x^2 \rangle$ $\ne$ $\langle u_z^2 \rangle$, $R_{k i}(0)$ $\ne$ 0, $i \ne k$)
and of $\nabla_{\bf x} {\bf U}$ leads to the following condition
\bea
\ds \frac{\partial U_i}{\partial x_k} R_{k i}(0) \ne 0
\label{aniso}
\eea
which can correspond to an energy production, where the energy spectrum 
is not more a function of $\bfkappa^2$.

\bigskip

At this point, we want to determine an useful scalar equation for describing the
main properties of the velocity correlation.
We start decomposing $R_{i j}$,  $H$ and $\bf U$ into an even function of 
$r \equiv \vert {\bf r} \vert$ (here called spherical part),  plus the remaining term:
\bea
\begin{array}{l@{\hspace{+0.2cm}}l}
R_{i j} (r_x, r_y, r_z)= \hat{R}_{i j}(r) + \Delta R_{i j} (r_x, r_y, r_z) \\\\
H (r_x, r_y, r_z) = \hat{H}(r) + \Delta H (r_x, r_y, r_z) \\\\
{\bf U}' - {\bf U}  = \hat{\bf U}(r) + \Delta {\bf U} (r_x, r_y, r_z) 
\end{array}
\label{dec}
\eea 
being
\bea
\begin{array}{l@{\hspace{+0.99cm}}l}
\ds \hat{R}_{i j}(r) =  
\frac{1}{6} \left( R_{i j}(r, 0, 0) + R_{i j}(0, r, 0)+ R_{i j}(0, 0, r) \right)  \\\\
\ds + \frac{1}{6}  \left( R_{i j}(-r, 0, 0) + R_{i j}(0, -r, 0)+ R_{i j}(0, 0, -r) \right), \\\\
\ds \hat{H}(r) =  \frac{1}{6} \left( H(r, 0, 0) + H(0, r, 0)+ H(0, 0, r) \right) \\\\ 
\ds + \frac{1}{6} \left( H(-r, 0, 0) + H(0, -r, 0)+ H(0, 0, -r) \right), \\\\
\ds \hat{\bf U}(r) =  \frac{1}{6} \left( {\bf U}(r, 0, 0) 
+ {\bf U}(0, r, 0)+ {\bf U}(0, 0, r) \right)  \\\\
\ds + \frac{1}{6} \left( {\bf U}(-r, 0, 0) + {\bf U}(0, -r, 0)+ 
{\bf U}(0, 0, -r) \right)
\end{array}
\label{isotrp}
\eea
where $\Delta R_{i j}({\bf 0})$=$\Delta H({\bf 0})$ = 0.
Therefore, the Fourier transform of $\hat{R}$ identifies the part of the energy spectrum depending upon $\bfkappa^2$ whose integral over the Fourier space gives the whole turbulent kinetic energy.
Moreover, $\hat{\bf U} \equiv 0$ and $\Delta {\bf U} \equiv {\bf U}'-{\bf U}$.

The laplacian of $R$ appearing into Eq. (\ref{cc2}) is written taking into account
 Eqs. (\ref{dec})  
\bea
\nabla^2 R = \frac{\partial^2 \hat{R}} {\partial r^2} +
\ds \frac{2}{r} \frac{\partial \hat{R}}{\partial r} +  \nabla^2 \Delta R  
\label{nabla2 R}
\eea
Substituting Eqs. (\ref{dec}), (\ref{isotrp}), (\ref{nabla2 R}) into Eq. (\ref{cc2}), we obtain the following equation
\bea
\ds
\frac{\partial \hat{R}}{\partial t} +  \frac{\partial \Delta R}{\partial t}  =
\frac{1}{2} \hat {H}  + \frac{1}{2} \Delta H 
+ 2 \nu \left(  \frac{\partial^2 \hat{R}} {\partial r^2} +
\ds \frac{2}{r} \frac{\partial \hat{R}}{\partial r}  \right)  
+ 2 \nu \nabla^2 \Delta R 
- \hat{G} - \Delta{G}
\label{cc3}
\eea
where
\bea
\begin{array}{l@{\hspace{+0.2cm}}l}
\ds \hat{G} = \frac{\partial U_i} {\partial x_k} \hat{R}_{k i}+ \hat{G}_0, \ \ \ \
\ds \Delta{G} = \frac{\partial U_i} {\partial x_k} \Delta{R}_{k i} 
- \Delta \left( \frac{\partial R }{\partial r_k} (U_k - U_k')\right) 
\end{array}
\eea
where $\hat{G}_0$ represents the spherical part of $-\partial R /\partial r_k (U_k - U_k') $.
As $\hat{R}$, $\hat {H}$ and $\hat{G}$ are even functions of $r$, whereas  $\Delta{R}$, $\Delta{H}$ and $\Delta{G}$ are not, and since Eq. (\ref{cc3}) must be satisfied at each time and for all the values of $\bf r$, Eq. (\ref{cc3}) is satisfied when 
\bea
\begin{array}{l@{\hspace{+0.2cm}}l}
\ds \frac{\partial \hat{R}}{\partial t} =
\frac{\hat{H}}{2} 
+ 2 \nu \left(  \frac{\partial^2 \hat{R}} {\partial r^2} +
\ds \frac{2}{r} \frac{\partial \hat{R}}{\partial r}  \right)  
- \hat{G}
\end{array}
\label{eq iso}
\eea
\bea
\begin{array}{l@{\hspace{+0.2cm}}l}
\ds \frac{\partial \Delta R}{\partial t}  =
\frac{ \Delta H}{2}
+ 2 \nu \nabla^2 \Delta R 
- \Delta{G}
\end{array}
\label{eq non-iso}
\eea
Thanks to decomposition (\ref{dec}), Eq.  (\ref{eq iso}) is about independent from 
Eq. (\ref{eq non-iso}), thus $\hat{R}$ is governed by a decoupled 
scalar equation which individually describes some of the properties of the  
non-isotropic turbulence related to $\hat{R}(r)$.
As far as Eq. (\ref{eq non-iso}) is concerned, $\Delta R$
%$\Delta R (t, r_x, r_y, r_z)$
is a complicate function of $\bf r$ and $t$ which depends on the particular problem.

Here, the turbulence is studied using Eq. (\ref{eq iso}) 
alone, whereas $R_{i j}({\bf r})$  will be determined in function of 
$\nabla_{\bf x}{\bf U}$, by means of a proper statistical analysis 
of the two-points velocity correlation.

\bigskip

\section{\bf Two-points distribution function}

%In order to analyse the non-isotropy caused by $\nabla_{\bf x} \bf U$ and
%to find a reasonable expression of $\hat{G}$, 
%the two-points distribution function of velocities and coordinates is 
%introduced. 
%To this purpose, consider the equations of motion of the various continuum
% fluid particles  

In order to obtain the pair distribution function, consider the equations of motion of the various continuum fluid particles  
\bea
\ds \frac{d{\bf x}_k}{dt}  = {\bf v}({{\bf x}_k}, t) \equiv {\bf v}_k, \ \  k=1, 2, ...N
\label{1}
\eea
This is an ordinary differential system, where ${\bf v}({{\bf x}}, t)$ 
varies according to the Navier-Stokes equations, here written for each 
fluid particle in the following form
\bea
\ds \frac{d {\bf v}_k}{dt} = \dot{\bf v}_k ({\bf v}_1, {\bf v}_2, ...,{\bf v}_k, ..., {\bf v}_N  ), \ k = 1, 2, ..., N
\label{NS}
\eea
being $\dot{\bf v}_k$ the acceleration of the $k^{th}$ particle.
This form of the Navier-Stokes equations is obtained once the pressure is eliminated through the continuity equation.

The distribution function of the velocities and of the spatial coordinates of all the fluid particles
\bea
F = F (t, {\bf x}_1, {\bf x}_2,...{\bf x}_N,{\bf v}_1,{\bf v}_2,...{\bf v}_N)
\label{F}
\eea
is defined in the phase space $\Gamma \equiv \Gamma_1 \times \Gamma_2 \times...\times \Gamma_N$, where 
$\Gamma_k = \left\lbrace  {\bf x}_k \right\rbrace \times  \left\lbrace {\bf v}_k \right\rbrace$, $k=1, 2, ...N$, being 
$\left\lbrace {\bf x}_k \right\rbrace$ = 
$\left\lbrace (-\infty, \infty) \times (-\infty, \infty) \times (-\infty, \infty) \right\rbrace$ 
and 
$\left\lbrace {\bf v}_k \right\rbrace$ =
$\left\lbrace (-\infty, \infty) \times (-\infty, \infty) \times (-\infty, \infty) \right\rbrace $
the subspaces of coordinates and velocities described by the $k^{th}$ particle.
This function, which vanishes on the boundaries of 
$\left\lbrace {\bf v}_k \right\rbrace$, ($k=1, 2, ...N$), 
satisfies the Liouville theorem associated to Eqs. (\ref{1}) and (\ref{NS}) (\cite{Nicolis95})
\bea
\ds \frac{\partial F }{\partial t} 
+ \sum_{k=1}^N \nabla_{{\bf x}_k}  F \cdot {\bf v}_k
= - \sum_{k=1}^N \nabla_{{\bf v}_k}  F \cdot \dot{\bf v}_k
\label{Liouville}
\eea
%where the RHS gives the rate of variation of $F$ due to $\dot{\bf v}_k$.

To study the correlation between two points of space
${\bf x} \equiv {\bf x}_1$ and ${\bf x}' \equiv {\bf x}_2$, 
the two-points distribution function $F^{(2)}$ associated to  
${\bf x}$ and ${\bf x}'$, is considered. 
This is the reduced distribution function calculated, by definition, integrating $F$ over  $\Gamma_3 \times \Gamma_4 \times ... \times \Gamma_N$ 
\bea
F^{(2)} = \int_{\Gamma_3} \int_{\Gamma_4} ... \int_{\Gamma_N} F \prod_{k =3}^N  d x^3_k dv^3_k  
\label{F^{(2)}}
\eea
Hence, $F^{(2)}$ obeys to the equation arising from Eq. (\ref{Liouville})
\bea
\ds \frac{\partial F^{(2)} }{\partial t}
+ \nabla_{{\bf x}}  F^{(2)} \cdot {\bf v}
+ \nabla_{{\bf x}'}  F^{(2)} \cdot {\bf v}'
= J 
\label{2-points}
\eea
where
\bea
J= -  \int_{\Gamma_3} \int_{\Gamma_4} ... \int_{\Gamma_N} 
\left( \sum_{k=3}^N \nabla_{{\bf x}_k} F \cdot {\bf v}_k 
+ \sum_{k=1}^N \nabla_{{\bf v}_k} F \cdot \dot{\bf v}_k \right) 
\prod_{k =3}^N  d x^3_k dv^3_k
\label{Jeq} 
\eea
is the rate of $F^{(2)}$ caused by the interactions between the two particles that simultaneously pass through $\bf x$ and $\bf x'$, and all the other fluid particles.

The local statistical equilibrium for the system of two fluid particles 
is expressed by the condition
\bea
J = 0
\label{stat eq}
\eea
To determine the expression of $J$, the case with null 
rate of kinetic energy is first considered.
We assume that the local statistical equilibrium corresponds
to the condition of fully developed homogeneous isotropic turbulence 
in an infinite fluid domain whose values of  
momentum and kinetic energy coincide with those of the current condition
\bea
\begin{array}{l@{\hspace{+0.2cm}}l}
\ds \int_{v} \int_{v'}  F^{(2)} du^3 du'^3= 
\ds \int_{v} \int_{v'}  F^{(2)}_0  du^3 du'^3, \\\\
\ds \int_{v} \int_{v'}  F^{(2)} {\bf u}  \ du^3 du'^3= 
\ds \int_{v} \int_{v'}  F^{(2)}_0 {\bf u} \ du^3 du'^3, \\\\
\ds \int_{v} \int_{v'}  F^{(2)} {\bf u} \cdot {\bf u} \ du^3 du'^3= 
\ds \int_{v} \int_{v'}  F^{(2)}_0 {\bf u}\cdot {\bf u} \  du^3 du'^3
\end{array}
\label{cons}
\eea
That is, $F^{(2)}_0$ represents the fully developed isotropic turbulence and
is related to $F^{(2)}$ through Eqs. (\ref{cons}).
Taking into account the homogeneity, the centered moments
of $F^{(2)}_0$ are constants in space, thus $F^{(2)}_0$ is a function of
${\bf v} -{\bf U}$ and ${\bf v}' -{\bf U}'$ 
\bea
F^{(2)}_0({\bf v}, {\bf v}'; {\bf x}, {\bf x}') =  F^{(2)}_0({\bf v} -{\bf U}({\bf x}), {\bf v}' -{\bf U}({\bf x}'))
\label{iso}
\eea

Now, $J$ is the rate of $F^{(2)}({\bf u}, {\bf u}')$ whose variations are 
caused by the fluctuations ($\tilde{\bf u}$, $\tilde{\bf u}'$)
\bea
\ds J \equiv \left( \frac{d F^{(2)}} {dt }\right)_{\bf \tilde{\bf u}, \tilde{\bf u}'}
\label{J}
\eea
These variations are determined using the finite scale Lyapunov analysis 
presented in \cite{deDivitiis_1}, where for sake of convenience, $J$ is calculated
for $\bf \tilde{\bf u}$=0, 
$\ds \bf \tilde{\bf u}' = \tilde{\bf u}'_0 {\mbox e}^{\lambda(r) t}$.
According to the theory, 
$\vert {\bf \tilde{\bf u}'_0} \vert << \vert {\bf u}' \vert$, and the variations of $F^{(2)}$ are calculated through $F^{(2)}_0$ and the Frobenius-Perron equation (\cite{Nicolis95})
\bea
\begin{array}{l@{\hspace{+0.2cm}}l}
\ds F^{(2)}({\bf u}, {\bf u}'+\tilde{\bf u}', t) =
\int_{-\infty}^{\infty} \int_{-\infty}^{\infty} \int_{-\infty}^{\infty}  F^{(2)}_0({\bf u}, {\bf u}'+\tilde{\bf u}'_0) 
\delta( \tilde{\bf u}'-\tilde{\bf u}'_0 {\mbox e}^{\lambda(r) t} ) d^3 u' \\\\
= F^{(2)}_0({\bf u}, {\bf u}'+\tilde{\bf u}' {\mbox e}^{-\lambda(r) t})
\end{array}
\label{F_P}
\eea
where $\delta$ is the Dirac delta and $\lambda(r) > 0$ is the maximal Lyapunov exponent of finite scale (\cite{deDivitiis_1}).
Being $\vert \tilde{\bf u}' {\mbox e}^{-\lambda(r) t} \vert << \vert {\bf u}' \vert$
\bea
\ds F^{(2)}_0({\bf u}, {\bf u}'+\tilde {\bf u}' {\mbox e}^{-\lambda(r) t}) \simeq
\ds F^{(2)}_0({\bf u}, {\bf u}') + \frac{\partial F^{(2)}_0}{\partial {\bf u}'}  \cdot  \tilde{\bf u}' {\mbox e}^{-\lambda(r) t} 
\label{ap}
\eea
and, taking into account Eqs. (\ref{F_P}) and (\ref{ap}), 
$J = \lambda(r) (F^{(2)}_0 -  F^{(2)})$.

\bigskip

In presence of a nonzero rate of kinetic energy, $J$ reads as 
\bea
\ds J = \lambda(r) \left(  F^{(2)}_0 -  F^{(2)}  \right) -J_D   
\label{J2}
\eea
where 
%$F^{(2)}$ and $F^{(2)}_0$ also exhibit the same energy rate of
% dissipation, and
 $-J_D$ is the rate of $F^{(2)}$ due to the rate of kinetic energy.
Now, the statistical equilibrium ($J=0$) differs from the isotropic turbulence, 
being $J_D \ne$0 into Eq. (\ref{J2}).

Therefore, the evolution equation of $F^{(2)}$ is assumed to be
\bea
\ds \frac{\partial F^{(2)} }{\partial t}  
+ \nabla_{{\bf x}}  F^{(2)} \cdot {\bf v} 
+ \nabla_{{\bf x}'}  F^{(2)} \cdot {\bf v}'  
= \lambda(r) \left(  F^{(2)}_0 -  F^{(2)}  \right) 
 -J_D   
\label{3}
\eea 
Accordingly, $F^{(2)}$ varies depending on the boundary conditions associated 
to Eq. (\ref{3}) and on the initial condition.
Equation (\ref{3}) is a partial differential equation, where $1/\lambda$ identifies the relaxation time of the system given by a pair of fluid particles.

\bigskip

\section{\bf  Effect of the average velocity gradient}

To study the effect of $\nabla_{\bf x} {\bf U}$,  consider now the homogeneous turbulence in a steady velocity gradient, whose pair distribution function is
\bea
F^{(2)} ({\bf v}, {\bf v}'; {\bf x}, {\bf x}') =  F^{(2)}_0 ({\bf v} -{\bf U} ({\bf x}), {\bf v}'-{\bf U} ({\bf x}')) + \phi^{(2)} ({\bf v}, {\bf v}'; {\bf x}, {\bf x}')
\label{iso0}
\eea
where $\phi^{(2)}$, representing the deviation from the isotropic turbulence,
satisfies, at each instant, Eqs. (\ref{cons})
\bea
\begin{array}{l@{\hspace{+0.2cm}}l}
\ds \int_{v} \int_{v'} \phi^{(2)} du^3 du'^3= 0, \\\\
\ds \int_{v} \int_{v'} \phi^{(2)} {\bf u}  \ du^3 du'^3= 0, \\\\
\ds \int_{v} \int_{v'}  \phi^{(2)} {\bf u} \cdot {\bf u} \ du^3 du'^3= 0.
\end{array}
\label{cons1}
\eea
and Eq. (\ref{3})
\bea
\begin{array}{l@{\hspace{+0.2cm}}l}
\ds \lambda \phi^{(2)} = 
 -J_D   
\ds - \left(  \frac{\partial F^{(2)}_0 }{\partial t} + \frac{\partial \phi^{(2)} }{\partial t} 
+  \frac{\partial \phi^{(2)} }{\partial x_p} v_p 
+  \frac{\partial \phi^{(2)} }{\partial x_p'} v_p'
  \right) \\\\
 \hspace{+15 mm} 
\ds +\left(
\frac{\partial F^{(2)}_0}{\partial v_j}  v_p 
+ \frac{\partial F^{(2)}_0}{\partial v'_j}  v'_p 
\right)  \frac{\partial U_j}{\partial x_p} 
%\hspace{+15 mm}  
\end{array}
\label{fi_2}
\eea
where the spatial derivatives of $F^{(2)}_0$ are written in function of 
 $\nabla_{\bf x} {\bf U} = \nabla_{\bf x'} {\bf U}'$ by means of Eq. (\ref{iso})
\bea
\frac{\partial F^{(2)}_0}{\partial x_k} = 
- \frac{\partial F^{(2)}_0}{\partial v_j} \frac{\partial U_j}{\partial x_k},
 \ \ \ \
\frac{\partial F^{(2)}_0}{\partial x'_k} = 
-\frac{\partial F^{(2)}_0}{\partial v'_j} \frac{\partial U_j}{\partial x_k}
\label{4}
 \eea 

With reference to Eq. (\ref{fi_2}), in an infinitive fluid domain,
where $\nabla_{\bf x} {\bf U} = 0$, Eq. (\ref{3}) admits solutions 
$F^{(2)}$ representing the homogeneous turbulence that decays because of $J_D \ne 0$.
For $\nabla_{\bf x} {\bf U} \ne  0$, 
the last  term at the RHS of Eq. (\ref{fi_2})
identically satisfies Eqs. (\ref{cons1}), thus $\phi^{(2)}$ must also satisfy 
the following set of equations  
\bea
\int_v \int_{v'}
\left[\begin{array}{c}
\hspace{-0.0mm} \ds 1 \\\
\hspace{-0.0mm} \ds {\bf u}  \\\
\hspace{-0.0mm} \ds {\bf u} \cdot {\bf u} 
\end{array}\right]  
 \left( \frac{\partial F^{(2)} }{\partial t} 
+  \frac{\partial \phi^{(2)} }{\partial x_p} v_p
+  \frac{\partial \phi^{(2)} }{\partial x_p'} v_p'
  +J_D   \right) \ du^3 du'^3 \equiv 0
\label{c0}
\eea
A sufficient condition for these five equations is 
\bea
\frac{\partial F^{(2)} }{\partial t} 
+  \frac{\partial \phi^{(2)} }{\partial x_p} v_p
+  \frac{\partial \phi^{(2)} }{\partial x_p'} v_p'
  +J_D   \equiv 0
\label{sc}
\eea
Assuming that Eq. (\ref{sc}) is true, and taking into account Eq. (\ref{fi_2}),
$\phi^{(2)}$ is a linear function of the velocity gradient and $F^{(2)}_0$ 
\bea
\ds F^{(2)} = F^{(2)}_0 + \frac{1}{\lambda(r)}
\left(
\frac{\partial F^{(2)}_0}{\partial v_j}  \frac{\partial U_j}{\partial x_p} v_p 
+ \frac{\partial F^{(2)}_0}{\partial v'_j}  \frac{\partial U_j}{\partial x_p} v'_p 
\right) 
\label{F2}
\eea
Of course, $F^{(2)}$ really changes starting from an arbitrary initial condition, therefore
Eq. (\ref{F2}) represents an approximation which can be considered to be valid far from the initial condition.

At this stage of the analysis, the tensor $R_{k i}$ is calculated, by definition 
\bea
\begin{array}{c@{\hspace{+0.2cm}}l}
\ds R_{k i} = \int_v \int_{v'} F^{(2)} u_k u'_i \  d^3u \ d^3u' =  R_{k i 0} \\\\
\ds + \frac{1}{\lambda} 
 \frac{\partial U_j}{\partial x_p}  \int_v \int_{v'} \left(
\frac{\partial}{\partial v_j} \left(  F^{(2)}_0 v_p u_k u_i' \right) 
- F^{(2)}_0 \frac{\partial}{\partial v_j} \left( v_p u_k u_i' \right) 
 \right) \  d^3u \ d^3u' \\\\
\ds + \frac{1}{\lambda} 
 \frac{\partial U_j}{\partial x_p} \int_v \int_{v'}
\left( \frac{\partial}{\partial v'_j} \left(  F^{(2)}_0 v_p' u_k u_i' \right) 
- F^{(2)}_0 \frac{\partial}{\partial v'_j} \left( v_p' u_k u_i' \right) 
 \right) \  d^3u \ d^3u'
\end{array}
\label{R_0}
\eea
Into Eq. (\ref{R_0}), the integrals
of ${\partial}/{\partial v_j} (  F^{(2)}_0 v_p u_k u_i' )$ 
and of ${\partial}/{\partial v'_j} (  F^{(2)}_0 v_p' u_k u_i' )$ 
are both equal to zero as they are the integrals of
$ F^{(2)}_0 v_p u_k u_i'$ and  $ F^{(2)}_0 v_p' u_k u_i'$ 
calculated over the boundaries of 
$\left\lbrace {\bf v} \right\rbrace$ and $\left\lbrace {\bf v}' \right\rbrace$, 
where $F_0^{(2)}$ identically vanishes.
Furthermore, taking into account that ${\bf U}$ is solenoidal, $R_{k i}$ is
\bea
R_{k i} = R_{k i 0} - \frac{1}{\lambda} 
\left( \frac{\partial U_k}{\partial x_p} R_{p i 0} 
+  \frac{\partial U_i}{\partial x_q} R_{k q 0} 
 \right) 
\label{R}
\eea
where 
$R_{k i 0}$ is the second order velocity correlation tensor for the isotropic turbulence  (\cite{Batchelor53})
\bea
R_{k i 0} ({\bf r}) = u^2 \left( (f -g) \frac{r_k r_i}{r^2} + g \delta_{k i} \right) \label{R0}
\eea
being $f$ and $g = f + 1/2 \ r \ \partial f/\partial r$ longitudinal and lateral
 velocity correlation functions, respectively.
Namely, according to the present analysis, $R_{k i}$ is the sum of the isotropic correlation tensor plus the further term due to $\nabla_{\bf x} {\bf U}$.

For $r = 0$, Eq. (\ref{R}) provides the expression of the Reynolds stresses
in function of $\nabla_{\bf x} {\bf U}$ 
\bea
\left\langle  u_k u_i \right\rangle = u^2 \left( \delta_{k i} 
-\frac{1}{\Lambda} \left( \frac{\partial U_k}{\partial x_i}   
+  \frac{\partial U_i}{\partial x_k}    \right)  \right) 
\label{Boussinesq}
\eea 
where $\Lambda = \lambda (0)$ is the maximal Lyapunov exponent.
Equation (\ref{Boussinesq}) coincides with the Boussinesq approximation, 
where the link between $\left\langle  u_k u_i \right\rangle$ and
 $\nabla_{\bf x} {\bf U}$ is represented by the ratio $u^2 / \Lambda$
which identifies the eddy viscosity.
In particular, the velocity standard deviations along $r_x$, $r_y$ and $r_z$ are
\bea
\begin{array}{c@{\hspace{+0.2cm}}l}
\ds \left\langle  u_x^2 \right\rangle = u^2 \left( 1  
-\frac{2}{\Lambda} \frac{\partial U_x}{\partial x}  \right), \\\\
\ds \left\langle  u_y^2 \right\rangle = u^2 \left( 1  
-\frac{2}{\Lambda} \frac{\partial U_y}{\partial y} \right), \\\\
\ds \left\langle  u_z^2 \right\rangle = u^2 \left( 1  
-\frac{2}{\Lambda} \frac{\partial U_z}{\partial z}  \right)
\end{array}
\eea
Therefore, $\ds \sum_{k=1}^3 \left\langle  u_k^2 \right\rangle = 3 u^2$.
According to the present analysis,  
$\nabla_{\bf x} {\bf U}$ does not modify the turbulent kinetic energy
with respect to the isotropic turbulence: It only generates a non-isotropic condition which distributes the kinetic energy $3/2 u^2$  along the three spatial directions in a different fashion, depending upon $\nabla_{\bf x} {\bf U}$.

\bigskip

Now, $\hat{R}_{k i}$ is calculated according to Eq. (\ref{isotrp})
\bea
\hat{R}_{k i} = \frac{u^2}{3} \left(  3 f + \frac{\partial f}{\partial r} r \right) \left( \delta_{k i} 
-\frac{1}{\lambda} \left( \frac{\partial U_k}{\partial x_i}   
+  \frac{\partial U_i}{\partial x_k}    \right)  \right) 
\label{R2}
\eea
and $\hat{R}$ and $\hat{G}$ are now expressed in view of Eq. (\ref{R2})
\bea
\begin{array}{c@{\hspace{+0.2cm}}l}
\ds \hat{R} = \frac{u^2}{2}\left( 3 f +\frac{\partial f}{\partial r} r \right),   \ \ \ \
\ds \hat{G} =
- \frac{S \ u^2}{3 \lambda(r)} \left( 3 f + \frac{\partial f}{\partial r} r \right) 
+ \hat{G}_0
\end{array}
\label{R3}
\eea
where $S$ is the frame invariant scalar 
\bea
S = \frac{\partial U_i}{\partial x_k}  
\left( \frac{\partial U_k}{\partial x_i}   
+  \frac{\partial U_i}{\partial x_k}  \right) \equiv
2 \left(  S_{i k} + Z_{i k} \right)   S_{i k} = 2 S_{i k} S_{i k} > 0
\eea
being
$
S_{k i} = 1/2 ( \partial U_k/\partial x_i + \partial U_i/\partial x_k)
$, 
$
Z_{k i} = 1/2 ( \partial U_k/\partial x_i - \partial U_i/\partial x_k)
$
and $-\hat{G} (0)>0$.
%Finally, 
%$\nabla_{\bf x} {\bf U}$ acts on $R_{i j}$, causing
% anisotropy
%according to Eq. (\ref{R}), and $\hat{G}$ corresponds to
% an energy production term.

Following this analysis,  $\nabla_{\bf x} {\bf U}$ influences the steady 
value of $R_{k i}$, therefore it is at the origin of the non-isotropy 
as prescribed by Eq. (\ref{R}), and of the kinetic energy production.
Moreover, being $\hat{G} (r) \ne 0$, $\nabla_{\bf x} {\bf U}$  is also
partially responsible for the kinetic energy transfer and for the 
shape of the energy spectrum.

\bigskip

\section{\bf Effect of the Rotating Reference Frame}

Now, we remark that Eq. (\ref{R3}) can be considered still valid when 
$\Re$ is a rotating frame of reference with a constant angular velocity 
$\bfOmega \equiv ((\Omega_{i j}))$.
In this case, Eq. (\ref{N-S}) includes the Coriolis terms 
$- 2 \bfOmega {\bf v}$, $- 2 \bfOmega {\bf v}'$ and  the centrifugal accelerations
$-\bfOmega \bfOmega {\bf x}$, $-\bfOmega \bfOmega {\bf x}'$.
These latter cause an increment of the average pressure (\cite{Batchelor67}), 
but do not provide any contribution to Eq. (\ref{cc1}) because 
they do not depend upon the fluid velocity. 
The Coriolis accelerations modify the evolution equation of $R_{i j}$, but do not alter the analytical structure of Eq. (\ref{eq iso}).
In fact, due to incompressibility, again $P_{i i}({\bf r}) \equiv0$ (\cite{Batchelor53}).
Moreover, since $\hat{R}_{i j}$ is a symmetric tensor and an even function of $r$, 
whereas $\Omega_{i j}$ is antisymmetric, Eq. (\ref{eq iso}) maintains its structure.

As the result, 
the RHS of the Liouville equation (\ref{Liouville}) includes the additional terms
$
- 2  \nabla_{{\bf v}_k} \cdot \left( F \ {\bfOmega} {\bf v}_k  \right) 
-  \nabla_{{\bf v}_k}  F \cdot   \bfOmega 
\bfOmega {\bf x}_k 
$
whose integrals over $\Gamma_3 \times \Gamma_4 ...\times \Gamma_N$
are equal to zero for $k >2$ as they are proportional to the integrals of 
$
- 2   \left( F \ {\bfOmega} {\bf v}_k  \right) 
-  \ F  \bfOmega 
\bfOmega {\bf x}_k 
$
calculated over the boundary of $\Gamma_3 \times \Gamma_4 ...\times \Gamma_N$
where $F \equiv 0$. 
In this case, the evolution equation of $F^{(2)}$ includes the Coriolis terms and the centrifugal acceleration
\bea
\begin{array}{c@{\hspace{+0.2cm}}l}
\ds \frac{\partial F^{(2)} }{\partial t}  
+ \nabla_{{\bf x}}  F^{(2)} \cdot {\bf v} 
+ \nabla_{{\bf x}'}  F^{(2)} \cdot {\bf v}'  
- 2  \nabla_{\bf v} \cdot \left( F^{(2)} \ {\bfOmega} {\bf v}  \right) \\\\
-2 \nabla_{\bf v'} \cdot \left( F^{(2)} \ {\bfOmega} {\bf v}'  \right) 
- \nabla_{\bf v'} F^{(2)} \cdot \bfOmega  \bfOmega {\bf r} 
= \lambda(r) \left(  F^{(2)}_0 -  F^{(2)}  \right) -J_D
\end{array}
\label{3a}
\eea
In a flow with ${\bfOmega} \ne$ 0, we assume that there exists a wide fluid region with 
a constant $\nabla_{\bf x} {\bf U} \ne$ 0, where the turbulence is homogeneous.
This zone can correspond to a fluid region in proximity 
of the rotation axis of the frame (\cite{Swinney2010}).

Substituting the pair distribution function $F^{(2)} = F^{(2)}_0 + \phi^{(2)}$
in Eq. (\ref{3a}), we obtain  
\bea
\begin{array}{l@{\hspace{+0.2cm}}l}
\ds \lambda \phi^{(2)} = 
 -J_D   
\ds +\left(
\frac{\partial F^{(2)}_0}{\partial v_j}  v_p 
+ \frac{\partial F^{(2)}_0}{\partial v'_j}  v'_p 
\right)  \frac{\partial U_j}{\partial x_p} \\\\
\ds + 2 \frac{\partial}{\partial v_p} \left( F^{(2)} \Omega_{p q} v_q \right) 
\ds + 2 \frac{\partial}{\partial v'_p} \left( F^{(2)} \Omega_{p q} v'_q \right) 
+ \frac{\partial}{\partial v'_p} \left( F^{(2)} \Omega_{p s} \Omega_{s q}  r_q \right) \\\\
\ds - \left(  \frac{\partial F^{(2)}_0 }{\partial t} + \frac{\partial \phi^{(2)} }{\partial t} 
+  \frac{\partial \phi^{(2)} }{\partial x_p} v_p 
+  \frac{\partial \phi^{(2)} }{\partial x_p'} v_p'
  \right) 
\end{array}
\label{fi_2 1}
\eea
where $\phi^{(2)}$ represents the deviation from the isotropic turbulence.

As before, the term proportional to $\nabla_{\bf x} {\bf U}$ provides null contribution 
into Eq. (\ref{cons}), so also the other ones proportional to $\bfOmega$, as these
latter are surface integrals calculated over the boundary of the velocity space.
Therefore Eq. (\ref{c0}) still holds in this case and Eq. (\ref{sc}) is again 
considered to be satisfied for solving Eq. (\ref{c0}). 
Thus, in view of Eq. (\ref{fi_2 1}), $\phi^{(2)}$ is
the sum of a linear term of $\nabla_{\bf x} {\bf U}$ (as in Eq. (\ref{F2})) plus
another one depending on  $\bfOmega$ 
\bea
\begin{array}{l@{\hspace{+0.2cm}}l}
\ds \phi^{(2)} = 
\ds \frac{1}{\lambda} \left(
\frac{\partial F^{(2)}_0}{\partial v_j}  v_p 
+ \frac{\partial F^{(2)}_0}{\partial v'_j}  v'_p 
\right)  \frac{\partial U_j}{\partial x_p} 
+ \phi^{(2)}_\Omega
\end{array}
\label{fi_2 2}
\eea
where
\bea
\begin{array}{l@{\hspace{+0.2cm}}l}
\ds \phi^{(2)}_\Omega = 
\frac{\Omega_{p s} \Omega_{s q}  r_q}{\lambda}
\frac{\partial F^{(2)}}{\partial v'_p} + 
\ds \frac{2 \Omega_{p q}}{\lambda} \left( \frac{\partial}{\partial v_p} \left( F^{(2)} v_q \right)
\ds + \frac{\partial}{\partial v'_p} \left( F^{(2)} v'_q \right) \right) 
\end{array}
\label{F2a}
\eea
Note that, $\phi^{(2)}$ really varies starting from an arbitrary initial condition, therefore
Eq. (\ref{fi_2 2})-(\ref{F2a}) represents an approximations which can be considered to be valid far from the initial condition.

The increment of the velocity correlation tensor associated to $\phi^{(2)}_\Omega$ is
then calculated  
\bea
\begin{array}{c@{\hspace{+0.2cm}}l}
\ds R_{{\Omega} k i} = 
\int_v \int_{v'} \phi_{\Omega}^{(2)} u_k u'_i \  d^3u \ d^3u' =  
\frac{\Omega_{p s} \Omega_{s q} r_q}{\lambda}  \int_v \int_{v'} 
\frac{\partial F^{(2)}}{\partial v_p'} u_k u_i' \  d^3u \ d^3u' \\\\ 
\ds  + \frac{2 \Omega_{j p}}{\lambda} 
 \int_v \int_{v'} \left(
\frac{\partial}{\partial v_j} \left(  F^{(2)} v_p u_k u_i' \right) 
- F^{(2)} \frac{\partial}{\partial v_j} \left( v_p u_k u_i' \right) 
 \right) \  d^3u \ d^3u'  \\\\
\ds + \frac{2 \Omega_{j p}}{\lambda} 
 \int_v \int_{v'}
\left( \frac{\partial}{\partial v'_j} \left(  F^{(2)} v_p' u_k u_i' \right) 
- F^{(2)} \frac{\partial}{\partial v'_j} \left( v_p' u_k u_i' \right) 
 \right) \  d^3u \ d^3u'
\end{array}
\label{R_0a}
\eea
Many terms appearing at the RHS of Eq. (\ref{R_0a}) vanish for several reasons.
The first addend can be reduced to the sum of an integral calculated over the boundaries of $\left\lbrace \bf v\right\rbrace $ and 
$\left\lbrace \bf v' \right\rbrace $ plus the term proportional to the average fluctuating velocity, therefore it is identically equal to zero.
The integrals of ${\partial}/{\partial v_j} (  F^{(2)} v_p u_k u_i' )$ 
and of ${\partial}/{\partial v'_j} (  F^{(2)} v_p' u_k u_i' )$ 
are both equal to zero because they are the integrals of
$F^{(2)} v_p u_k u_i'$ and  $ F^{(2)} v_p' u_k u_i'$ 
calculated over the boundaries of $\left\lbrace {\bf v} \right\rbrace$ and 
$\left\lbrace {\bf v}' \right\rbrace$, where $F^{(2)} \equiv 0$. In conclusion, $R_{{\Omega} k i}$ is
\bea
R_{{\Omega} k i} =  - \frac{2 }{\lambda} 
\left(  \Omega_{k p} R_{p i} 
+   \Omega_{i q} R_{k q} 
 \right) 
\label{Ra}
\eea
Therefore
\bea
\ds \hat{R}_\Omega \equiv 0, \ \ \hat{G}=\frac{\partial U_i}{\partial x_k} 
\left(  \hat{R}_{k i} + \hat{R}_{\Omega k i} \right) + \hat{G}_0
\eea
Hence, taking into account Eq. (\ref{Ra})
\bea
\ds  \hat{G} = - \left( 3 f + \frac{\partial f}{\partial r} r\right) \frac{u^2}{3 \lambda} S
\left( 1 - \frac{2}{S \lambda} \frac{\partial U_i}{\partial x_k} \left( \Omega _{k p} S_{p i} + \Omega _{i p} S_{p k} \right)   \right)  + \hat{G}_0
\label{G rot}
\eea
Being $S_{i j}$ and $Z_{i j}$ symmetric and antisymmetric tensors, we obtain
\bea
\ds \frac{\partial U_i}{\partial x_k} \left( \Omega _{k p} S_{p i} + \Omega _{i p} S_{p k} \right) \equiv 0
\eea 
that is, $\partial U_i/\partial x_k \hat{R}_{\Omega k i} \equiv 0$ 
and Eqs. (\ref{R3}) and (\ref{eq iso}) are recovered.
This result is consistent with the fact that the Coriolis acceleration is
identically  normal to the fluid velocity, thus
this does not produce neither work nor energy transfer.

Observe that, although $\hat{R}$ does not depend on $\bfOmega$, 
the correlation tensor $R_{i j}$
is related to $\bfOmega$ through Eq. (\ref{Ra}).
%This result is consequence of the fact that the Coriolis
% force, acting on each fluid particle, is normal to the
% velocity, produces null work and does not determine any
% energy transfer.

\bigskip

\section{\bf Equation for the longitudinal correlation function}

Here, the evolution equation for the longitudinal correlation function
associated to $R_{i k 0}$, is derived.

Substituting Eqs. (\ref{R2}) and (\ref{R3}) into Eq. (\ref{eq iso}), we have
\bea
\begin{array}{c@{\hspace{+0.2cm}}l}
\ds   
\frac{\partial }{\partial t} \left( \frac{u^2}{2} \left( 3 f +\frac{\partial f}{\partial r} r \right) \right) =
\frac{\hat{H}}{2} 
+ 2 \nu \nabla^2 \hat{R} 
+ \frac{S \ u^2}{3} \left( 3 \frac{f}{\lambda} 
+   \frac{\partial}{\partial r} \left( \frac{f}{\lambda} \right) r  \right) \\\\
+ \ds \frac{S u^2}{3 \lambda^2} f \frac{\partial \lambda}{\partial r} r - \hat{G}_0
\end{array}
\label{c2}
\eea
First and second terms at the RHS give the mechanism of the energy cascade and the energy dissipation, respectively, whereas the third term, proportional to $S$, is greater than zero for $r=0$, thus it provides the turbulent energy production. 
The fourth one, also proportional to $S$, vanishes for $r = 0$ and does not contribute to the kinetic energy rate.

Observe that, into Eq. (\ref{c2}), $\hat{H}(r)$ is expressed without lack of generality as
\bea
\ds \hat{H}(r) = 3 K  + r \frac{\partial K}{\partial r}  + \varphi
\eea
where $K$, responsible for the energy cascade in isotropic turbulence, 
is related to the triple correlation function 
$
\ds k(r)= {\left\langle u_r^2({\bf x}) u_r({\bf x}+{\bf r}) \right\rangle}/{u^3}
$
through the relationship (\cite{Batchelor53})
\bea
K(r)= u^3 \left(  \frac{\partial}{\partial r}  + \frac{4}{r}  \right) k(r)
\label{kk}
\eea
whereas $\varphi$, representing the deviation from the isotropic turbulence, 
is an assigned function of $\nabla_{\bf x} {\bf U}$ and $f$.
According to the Lyapunov theory of finite scale proposed by 
\cite{deDivitiis_1}, $K$ and $\lambda$ are both in terms of $f$
\bea
\begin{array}{c@{\hspace{+0.2cm}}l}
\ds \lambda = \frac{u}{r} \sqrt{2(1-f)} \\\\
\ds K = u^3 \sqrt{\frac{1-f}{2}} \frac{\partial f}{\partial r}
\end{array}
\label{lyap}
\eea
As the result, Eq. (\ref{c2}) admits the following first integral
\bea
\begin{array}{c@{\hspace{+0.2cm}}l}
\ds \frac{\partial u^2 f}{\partial t}  = 
K  + 
\ds 2 \nu u^2 \left(  \frac{\partial^2 f} {\partial r^2} +
\ds \frac{4}{r} \frac{\partial f}{\partial r}  \right)  
+ \frac{2}{3}   \frac{S  u^2 f}{\lambda} \\\\
+ \ds \frac{1}{r^3}  \int_0^r \left( \frac{2}{3} S  u^2 \frac{f r^3}{\lambda^2} 
\frac{\partial \lambda}{\partial r}  +  r^2 \hat{\varphi} \right)  dr
\end{array}
\label{I1}  
\eea
being $\hat{\varphi} = \varphi - 2 \hat{G}_0$.
The first three terms appearing at the RHS of Eq. (\ref{I1}), identically satisfy the continuity equation (\cite{Batchelor53}) for an arbitrary correlation function
verifying the incompressibility condition.
From Eq. (\ref{lyap}) and assuming that the integral of $ r^2 \hat{\varphi}$ converges,
the last term of Eq. (\ref{I1}) behaves like $r^{-3}$ for $r \rightarrow \infty$. 
Therefore, the continuity equation is satisfied when 
\bea
\int_0^\infty \left( \frac{2}{3} S  u^2 \frac{f r^3}{\lambda^2} 
\frac{\partial \lambda}{\partial r}  +  r^2 \hat{\varphi} \right)  dr = 0
\label{continuity}
\eea
at each time and for arbitrary $f$ satisfying the continuity equation.
Being $\hat{\varphi}$ depending upon $f$ and $\nabla_{\bf x} {\bf U}$,
the integrand of Eq. (\ref{continuity}) identically vanishes and
$\hat{\varphi}$ is in terms of $f$ and $S$
\bea
\hat{\varphi}  = - \frac{2}{3} S  u^2 \frac{f r}{\lambda^2} 
\frac{\partial \lambda}{\partial r}  
\eea
\\
As the consequence, Eq. (\ref{I1}) becomes 
\bea
\ds \frac{\partial u^2 f}{\partial t}  = 
K   +
\ds 2 \nu u^2 \left(  \frac{\partial^2 f} {\partial r^2} +
\ds \frac{4}{r} \frac{\partial f}{\partial r}  \right)  
+ \frac{2}{3}   \frac{S  u^2 f}{\lambda} 
\label{I2}  
\eea
whose boundary conditions are  (\cite{Karman38},  \cite{Batchelor53}) 
\bea
f(0) = 1, \ \ \lim_{r \rightarrow \infty} f(r) = 0
\label{bc0}
\eea
For $\nabla_{\bf x} {\bf U} \rightarrow$0, Eq. (\ref{I2}) tends to   
the classical von K\'arm\'an-Howarth equation.

Putting $r=0$ into Eq. (\ref{I2}), we obtain the equation of evolution of the
turbulent kinetic energy
\bea
\ds \frac{d u^2}{dt} = 
\left( 10 \nu f''_0 + \frac{2}{3} \frac{S}{u \sqrt{-f''_0}} \right) u^2
\label{du^2/dt}
\eea
which states that
\bea
\begin{array}{c@{\hspace{+0.2cm}}l}
\ds \frac{S} {\Lambda^2} 
% \geq< 
\begin{array}{c@{\vspace{-1.5mm}}}
\ge \\
<
\end{array}
\frac{15}{R_T} \ \ \ \mbox{then} \ \  \frac{d u^2}{dt}  
\begin{array}{c@{\vspace{-1.5mm}}}
\ge \\
<
\end{array}
%\ge, < 
0,
  \ \ \ \mbox{respectively} 
\end{array}
\label{Ek1}
\eea
being $\lambda_T \equiv \sqrt{-1/f''_0}$ and $R_T = {u \lambda_T}/{\nu}$ the Taylor scale  and the Taylor scale Reynolds number, and $\Lambda = u /\lambda_T$.

The Fourier Transforms of $f$ and $K$ are (\cite{Batchelor53})
\bea
\left[\begin{array}{c}
\ds E(\kappa) \\\\
\ds T(\kappa)
\end{array}\right]  
= 
 \frac{1}{\pi} 
 \int_0^{\infty} 
\left[\begin{array}{c}
 \ds  u^2 f(r) \\\\
 \ds K(r)
\end{array}\right]  \kappa^2 r^2 
\left( \frac{\sin \kappa r }{\kappa r} - \cos \kappa r  \right) d r 
\label{Ek}
\eea
where $E(\kappa)$ and $T(\kappa)$ are the parts of the energy spectrum and of the so called transfer function which depend on $\bfkappa^2$.

\bigskip

This analysis states that the turbulence is described by the isotropic correlation tensor $R_{i k 0}$
(i.e.  $f(r)$), and by a non-isotropic term which  depends upon $\nabla_{\bf x} {\bf U}$.
The determination of $R_{i k}$ has been reduced into the following two steps:
\ \
1) Calculation of $f$ and $R_{i k 0}$ through Eq. (\ref{I2}) in function of $S$.
\ \
2) Calculation of $R_{i k}$ in terms of $\nabla_{\bf x} {\bf U}$ and $\bfOmega$ with Eqs. (\ref{R}) and (\ref{Ra}).

\bigskip

\section{\bf Steady Solutions}

According to Eq. (\ref{Ek1}), the steady correlation functions are obtained for
\bea
\frac{S} {\Lambda^2} = \frac{15}{R_T} 
\eea
with $\partial f / \partial t=0$, thus the dimensionless equation of $f$ is
\bea
\begin{array}{l@{\hspace{-0.cm}}l}
\ds  \sqrt{\frac{1-f}{2}} \ \frac{d f} {d \hat{r}}  +
\ds \frac{2}{R_T}  \left(  \frac{d^2 f} {d \hat{r}^2} +
\ds \frac{4}{\hat{r}} \frac{d f}{d \hat{r}}  \right) + \frac{10}{R_T} 
\frac{f \ \hat{r}}{\sqrt{2(1-f)}} = 0
\end{array}
\label{I2 steady}  
\eea
where $\hat{r} = r/\lambda_T$ and the boundary conditions are expressed by
 Eqs. (\ref{bc0}).
As the solutions $f$ $\in C^2$ $\left[0, \infty \right)$ tend to zero for $r\rightarrow\infty$, with  ${d^2 f (0)}/{d \hat{r}^2}=-1$, 
 the boundary conditions (\ref{bc0}) can be substituted with the condition
\bea 
\ds f(0)=1, \ \ \ \frac{d f (0)} {d \hat{r}} = 0
\label{bc3}
\eea
%\suppressfloats
 \begin{figure}[b]
	\centering
         \includegraphics[width=0.55\textwidth]{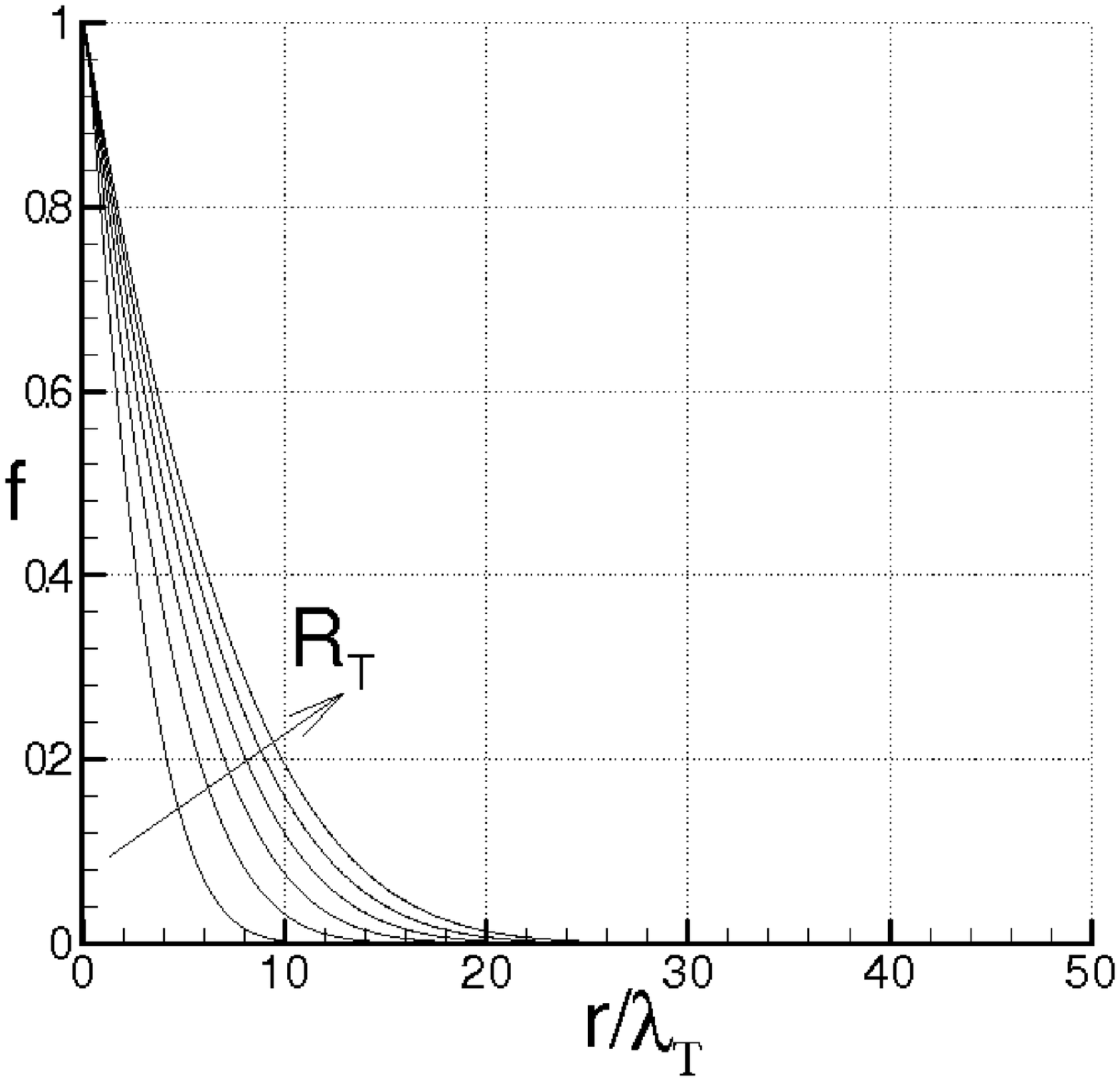}
\caption{Longitudinal correlation function for different Taylor-Scale Reynolds numbers.}
\label{figura_1}
\end{figure}
\begin{figure}[b]
	\centering         \includegraphics[width=0.55\textwidth]{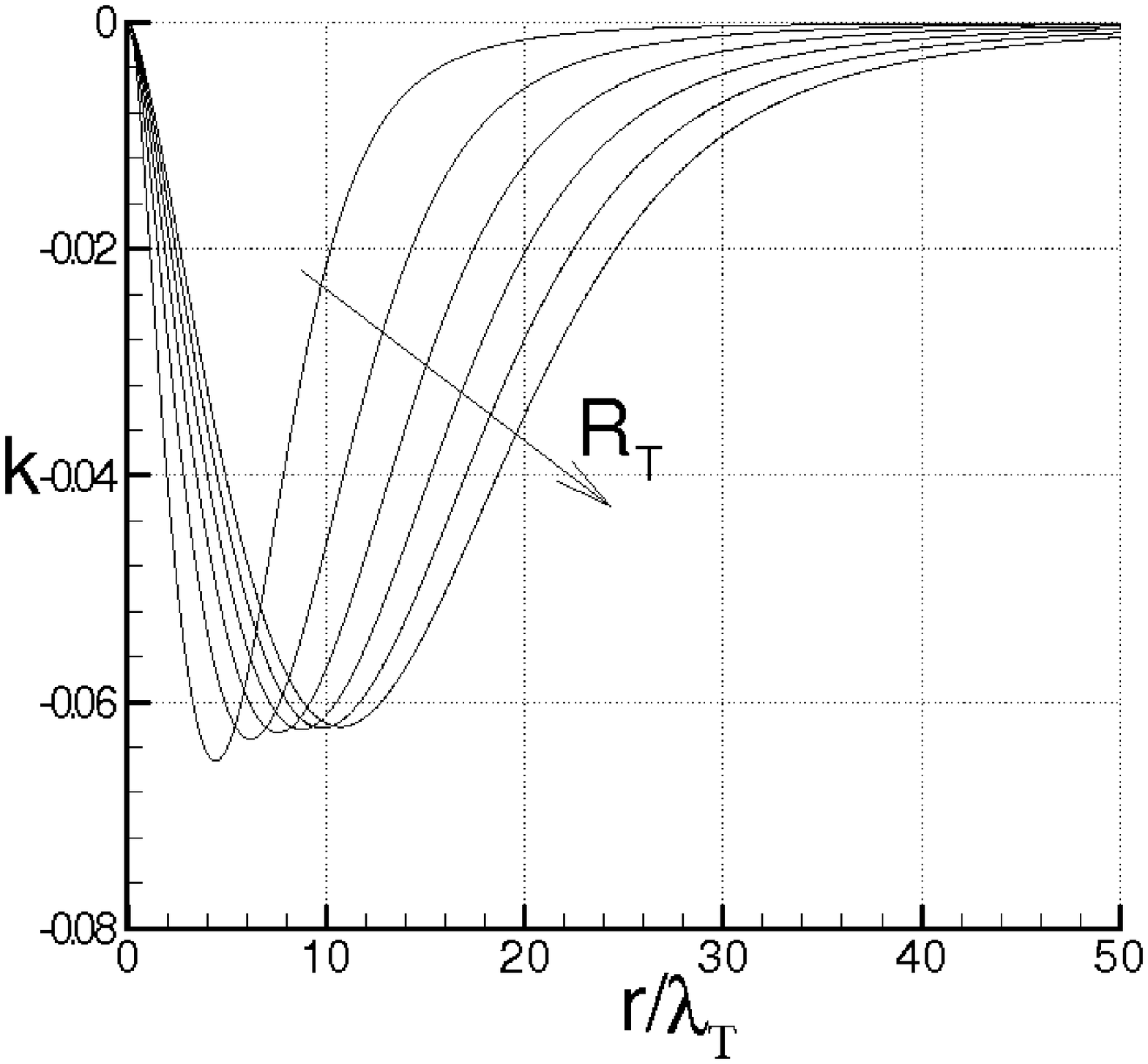}
\caption{Longitudinal triple correlation function for different Taylor-Scale Reynolds numbers.}
\label{figura_2}
\end{figure}
Therefore, the boundary problem represented by Eqs. (\ref{I2 steady}), (\ref{bc0}), is replaced by the following initial condition problem
\bea
\begin{array}{l@{\hspace{+0.cm}}l}
\ds  \frac{d f}{d \hat{r}} =  F \\\\
\ds \frac{d F}{d\hat{r}} = - \frac{5 f  \hat{r}}{\sqrt{2(1-f)}} -
\left( \frac{1}{2} \sqrt{\frac{1-f}{2}} R_T + \frac{4}{\hat{r}} \right) F
\end{array}
\label{vk-h2}  
\eea
whose initial condition is 
\bea 
\ds f(0) = 1, \ F(0) = 0
\label{ic}
\eea

\bigskip

\section{\bf Results and Discussion}

\suppressfloats
\begin{figure}[t]
	\centering
         \includegraphics[width=0.65\textwidth]{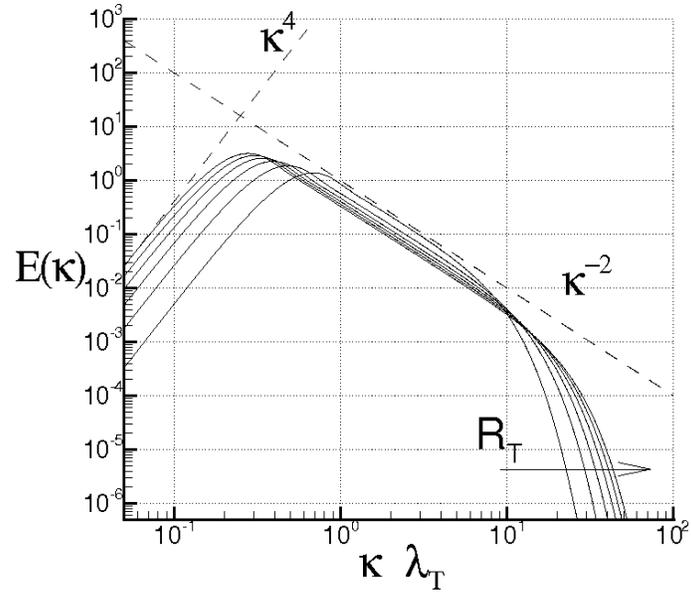}
\caption{Steady Energy Spectrum for different Taylor-Scale Reynolds numbers.}
\label{figura_3}
\end{figure}
\begin{figure}[t]
       \centering
        \includegraphics[width=0.65\textwidth]{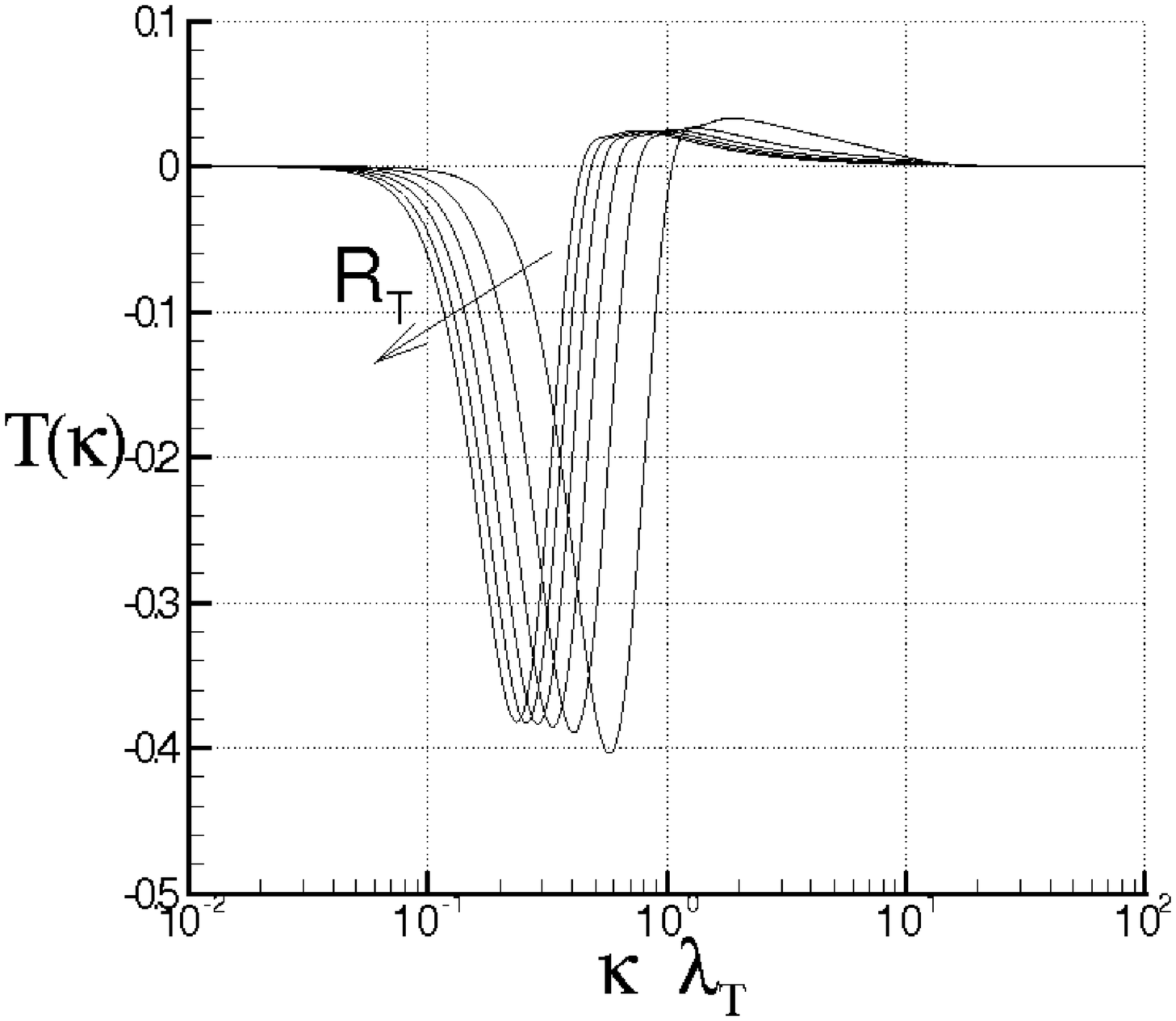}
\caption{Steady "Transfer function $T(\kappa)$" for several 
Taylor-Scale Reynolds numbers.}
\label{figura_4}
\end{figure}
It is worth to remark that Eq. (\ref{I2 steady}) is quite similar to the equation 
obtained by \cite{deDivitiis_2} in the case of self-similar isotropic  turbulence. 
\begin{figure}[t]
       \centering
        \includegraphics[width=0.65\textwidth]{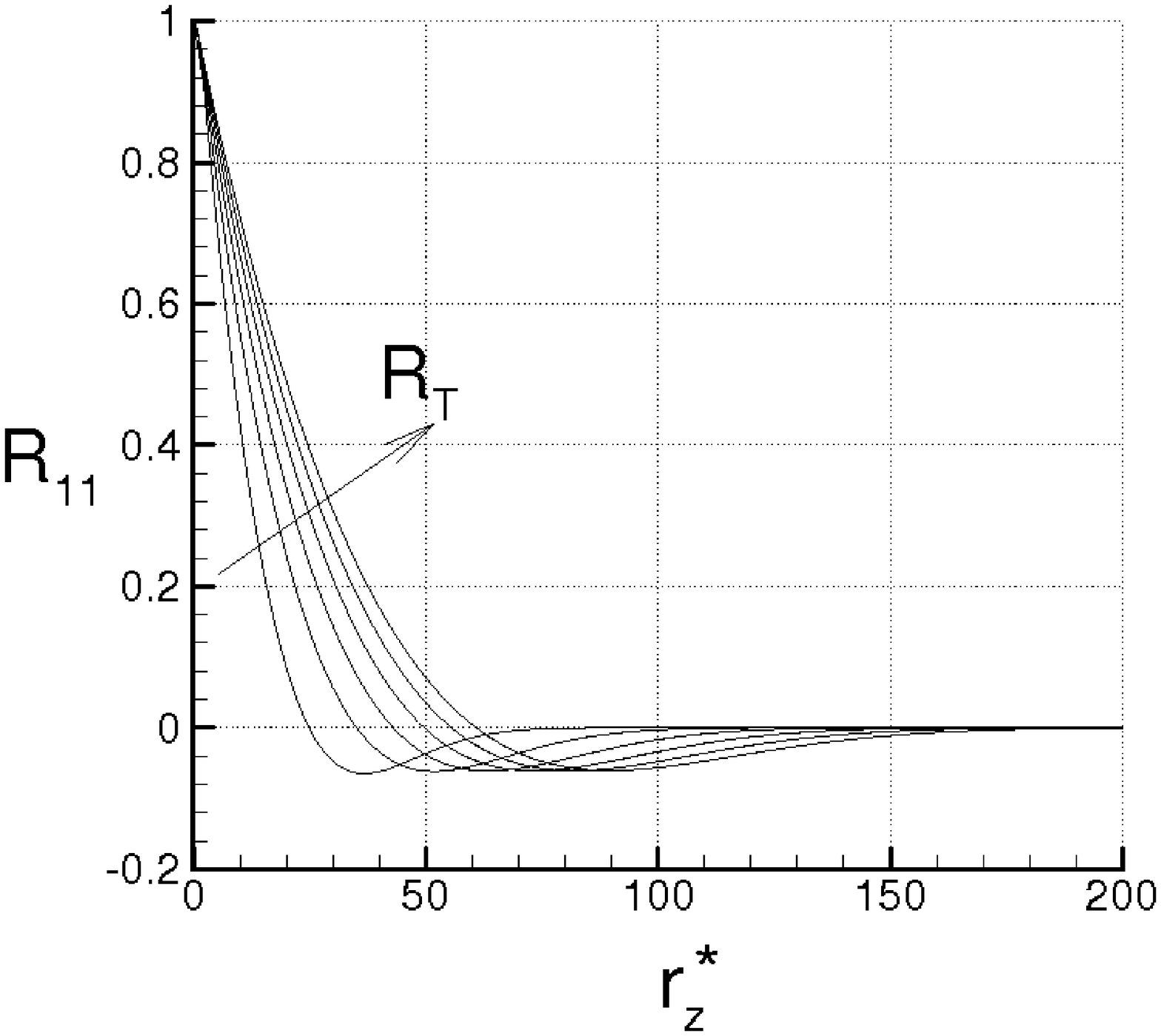}
\caption{Spanwise correlation function of the streamwise component of velocity for several
Taylor-Scale Reynolds numbers.}
\label{figura_1a}
\end{figure}
There, the equation was  
\bea
\begin{array}{l@{\hspace{-0.cm}}l}
\ds  \sqrt{\frac{1-f}{2}} \ \frac{d f} {d \hat{r}}  +
\ds \frac{2}{R_T}  \left(  \frac{d^2 f} {d \hat{r}^2} +
\ds \frac{4}{\hat{r}} \frac{d f}{d \hat{r}}  \right) + \frac{10}{R_T} f
= 0
\end{array}
\label{Isimilar}  
\eea
where again, ${d^2 f (0)}/{d \hat{r}^2}=-1$ and the boundary
conditions are expressed by Eq. (\ref{bc3}).
Equation (\ref{I2 steady}) differs from Eq. (\ref{Isimilar}) by the presence of the last term, that, into  Eq. (\ref{Isimilar}), arises from 
$\partial u^2 / \partial t$ under the self-similarity hypothesis.
\cite{deDivitiis_2} shows that $f -1 \approx r^{2/3}$, where 
$2/R_T  \left( d^2 f/ d \hat{r}^2 + 4/\hat{r} d f/d \hat{r} \right)$ 
is negligible with respect to the other terms (see Eq. (\ref{Isimilar})). 
This determines that $E(\kappa) \approx \kappa^{-5/3}$ 
in the inertial subrange and that the scaling law  
$\langle (\Delta u_r)^n \rangle \approx r^{\zeta_n}$  is satisfied for 
$\zeta_n \simeq n/3$.
Here, the presence of third term of Eq. (\ref{I2 steady}) 
modifies the correlation function, resulting  $f -1 \approx r$ and this implies 
that $E(\kappa) \approx \kappa^{-2}$ in the inertial subrange and that 
$\zeta_n \simeq n/2$.
\suppressfloats
\begin{figure}[t]
	\centering
         \includegraphics[width=0.65\textwidth]{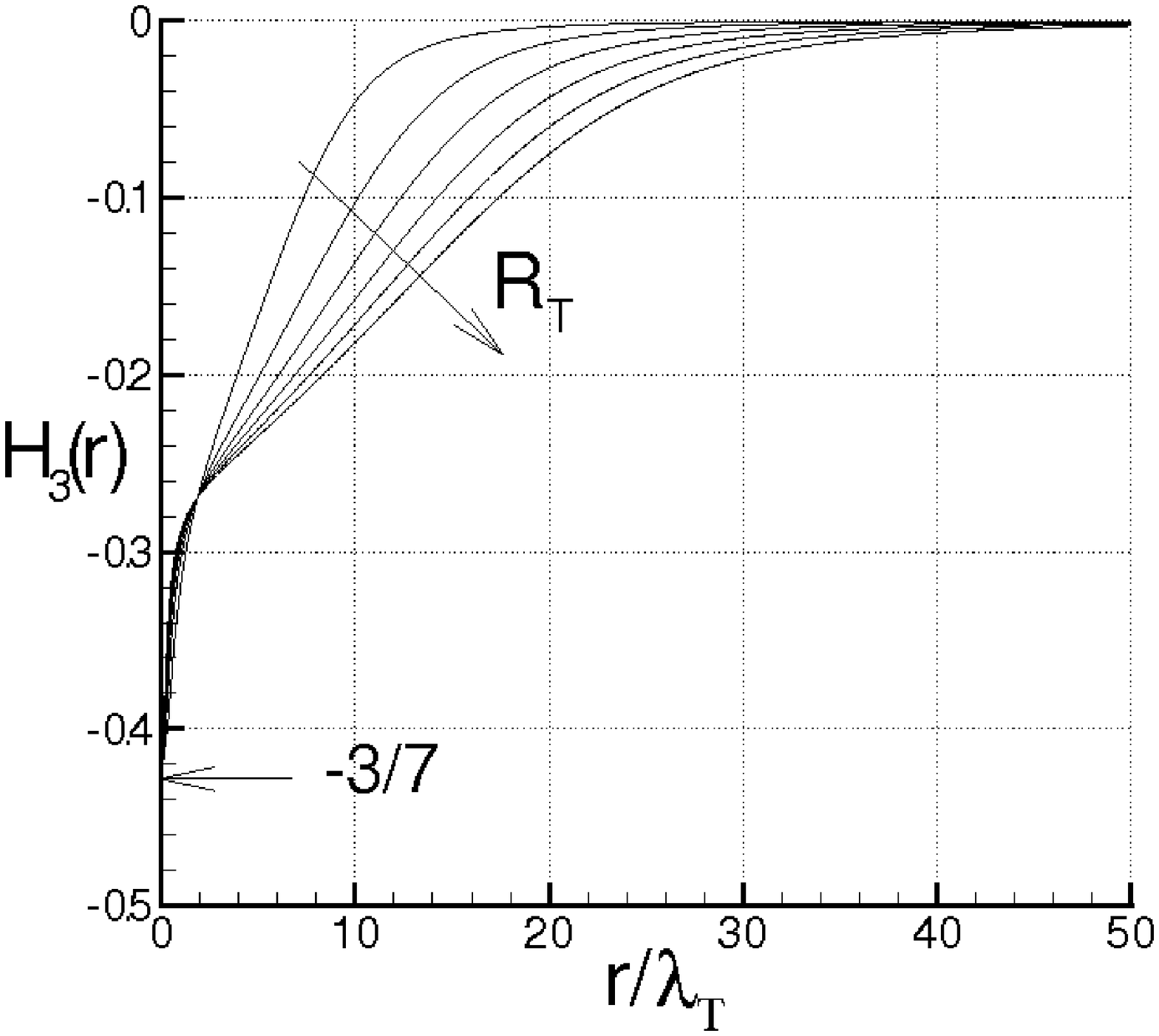}
\caption{Skewness of $\Delta u_r$ at different Taylor-Scale Reynolds numbers.}
\label{figura_5}
\end{figure}
\begin{figure}[t]
	\centering
         \includegraphics[width=0.65\textwidth]{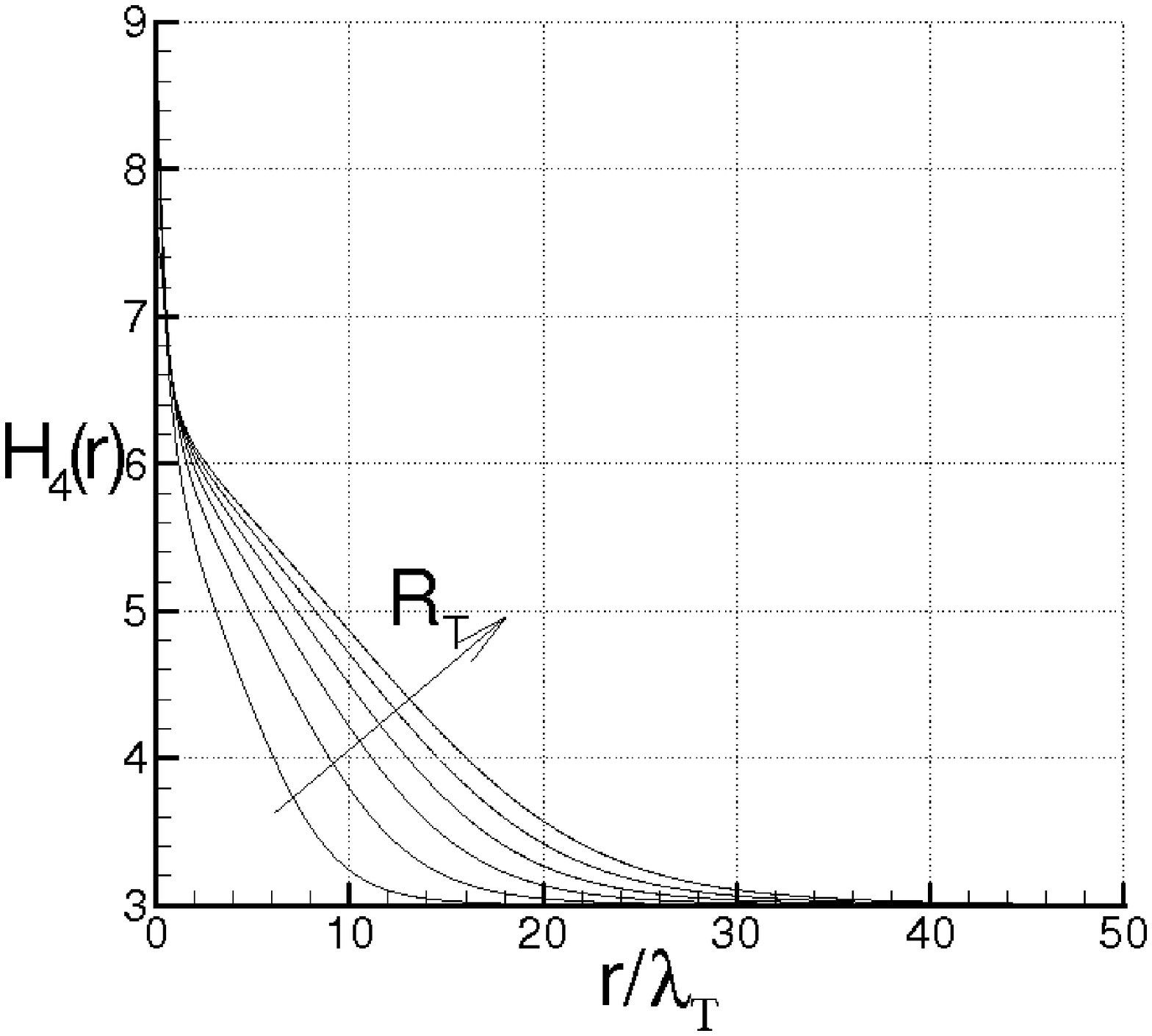}
\caption{Flatness of $\Delta u_r$ at different Taylor-Scale Reynolds numbers.}
\label{figura_6}
\end{figure}

To study in detail the energy spectrum, several numerical solutions of Eqs. (\ref{vk-h2}) are calculated for different values of $R_T$  by means of the fourth-order Runge-Kutta method.
The step of integration is chosen on the basis of the 
behavior of Eqs. (\ref{vk-h2}) at the small scales when $r \rightarrow \infty$.
There, $d^2 f / d\hat{r}^2$ is about equal to $K$
\bea
\begin{array}{l@{\hspace{-0.cm}}l}
\ds \frac{d^2 f} {d \hat{r}^2} \approx 
\ds -  \frac{R_T}{2 \sqrt{2}} \frac{d f} {d \hat{r}} 
\end{array}
\label{lvk}
\eea
This equation suggests that $\Delta \hat{r} = \sqrt{2}/R$ can be an adequate step of integration (\cite{NAG}) for the accuracy of the numerical solutions of  Eq. (\ref{vk-h2}).

As in \cite{deDivitiis_2}, the cases here studied correspond to 
$R_T$ = $100$, $200$, $300$, $400$, $500$ and $600$.

Figures \ref{figura_1} and \ref{figura_2} show double and triple longitudinal correlation functions. 
Due to the mechanism of energy cascade expressed by $K$ ( see Eq. (\ref{lyap})), 
the tail of $f$ rises with $R_T$ in agreement with Eq. (\ref{I1}) 
and this determines that, for an assigned value of 
$\lambda_T$, all the integral scales of $f$ increase with $R_T$.
Comparing these results with the corresponding data of \cite{deDivitiis_2},
we found that, the integral scales here calculated are about one-sixth those obtained in the case of homogeneous isotropic turbulence.
This means that $\nabla_{\bf x} {\bf U}$ contrasts the mechanism of the energy cascade, 
making this mechanism more limited in scales.

According to Eq. (\ref{lyap}), $k$ decays more slowly than $f$ and its characteristic scales increase with $R_T$ as prescribed by Eq. (\ref{vk-h2}).  
The maximum of $\vert k \vert$ is about 0.06 and this is in very good agreement with the numerous data of the literature (see \cite{Batchelor53} and Refs. therein).
Again, the scales of variation of $k$ are almost one-sixth those calculated in 
\cite{deDivitiis_2}.

Figures \ref{figura_3} and \ref{figura_4} show $E(\kappa)$ and $T(\kappa)$ calculated with Eq. (\ref{Ek}).
Because of the fluid incompressibility (see Eq. (\ref{continuity})), 
$E(\kappa) \approx \kappa^4$ near the origin, whereas $E(\kappa) \approx \kappa^{-2}$ in the inertial subrange in contrast with Kolmogorov law $\kappa^{-5/3}$. 
This disagreement, caused by $\nabla_{\bf x} {\bf U}$, makes
the mechanism of energy cascade weaker with respect to the isotropic turbulence, in agreement with the previous observation, determining a higher absolute slope of $E(\kappa)$ in the inertial range.
For what concerns $T(\kappa)$, as $K$ does not modify the fluid kinetic energy,  
$\int_0^{\infty} T(\kappa) d \kappa =$ 0 (see Fig. \ref{figura_4}).

\bigskip

Note that these cases can represent homogeneous turbulent flows with uniform steady shear rate
\bea 
\nabla_{\bf x} {\bf U} = \frac{\partial U} {\partial y}
\left[\begin{array}{ccc}
\ds 0 & 1 & 0 \\\\
\ds 0 & 0 & 0 \\\\ 
\ds 0 & 0 & 0
\end{array}\right]  
\eea
where $x$ and $z$ are streamwise and spanwise coordinates, respectively.
In this situation we expect that $\nabla_{\bf x} {\bf U}$ leads to the development of coherent structures, similar to streaks, caused by the stretching of the vortex lines along $x$  (\cite{Kim}).
This influences the spanwise correlation function of the streamwise velocity $R_{1 1} (r_z)$ which is here calculated through Eq. (\ref{R}), once $f$ is known.
The results, shown in Fig. \ref{figura_1a}, give $R_{1 1} (r_z)$ in terms of the dimensionless
coordinate $r_z^* = r_z / (\nu/\sqrt{S})^{1/2}$ and show that $R_{1 1} (r_z)$
intersects the horizontal axis and remains negative for $r_z \rightarrow \infty$.
These negative values imply a wide distribution 
of spacings between the different streaky structures, whose mean  
value depends on $R_T$ (\cite{Kim}). From the figure, this average distance results to be about 
$70 \div 150$ in units of viscous scale $(\nu/\sqrt{S})^{1/2}$, comparable with the results of 
Ref. (\cite{Kim}).

\bigskip

Figures \ref{figura_5} and \ref{figura_6} illustrate skewness and flatness 
of $\Delta u_r$ for the same values of $R_T$.
The skewness $H_3$ is first calculated with Eq. (\ref{H_3_01}) and the flatness $H_4$ has been thereafter determined using Eq. (\ref{m1}).
\suppressfloats
\begin{figure}[t]
	\centering
         \includegraphics[width=0.65\textwidth]{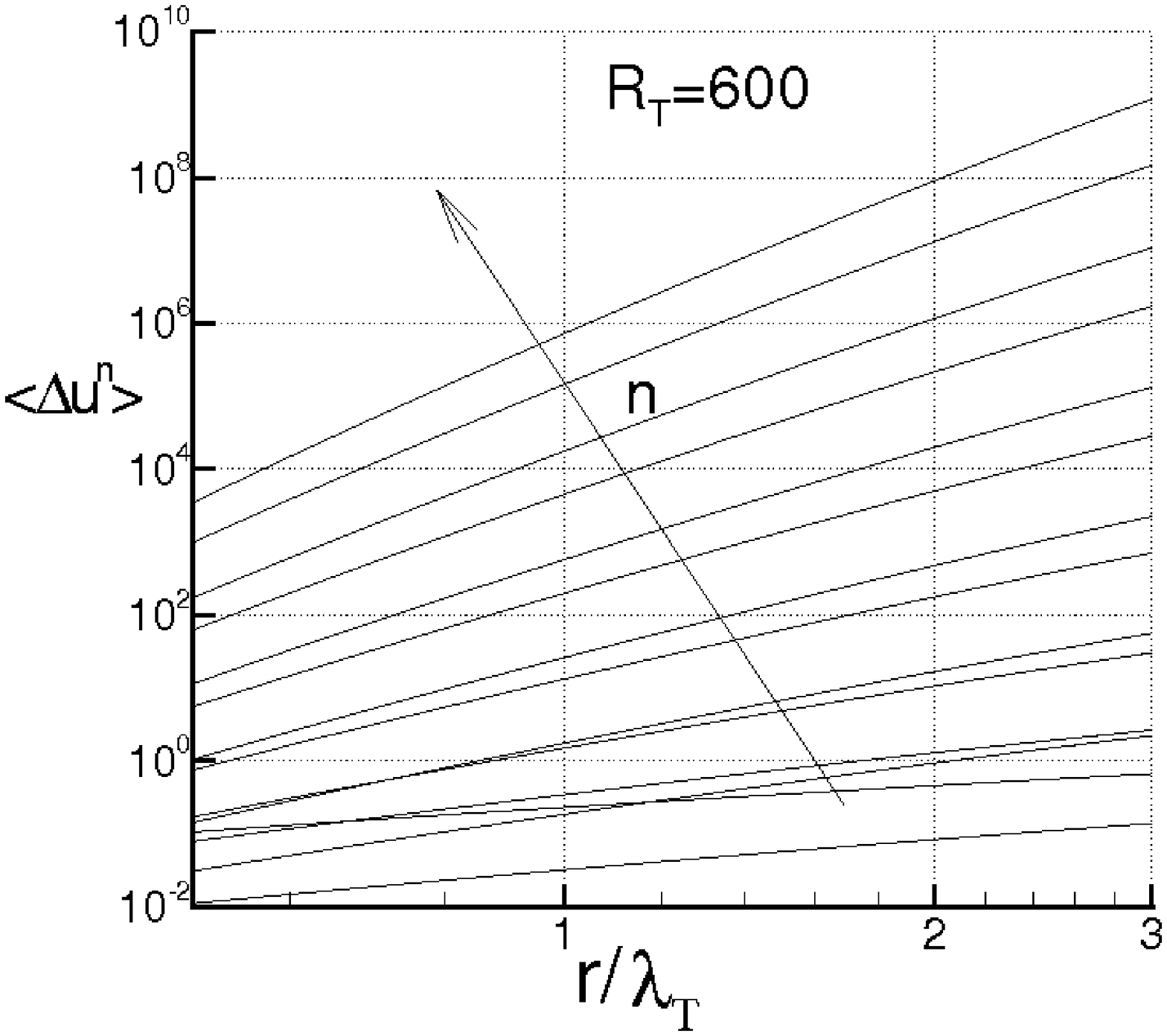}
\caption{Statistical moments of $\Delta u_r$ in terms of
the separation distance, for $R_T$=600.}
\label{figura_8}
\end{figure}
\suppressfloats
\begin{figure}[t]
%\vspace{-0.mm}
	\centering
         \includegraphics[width=0.65\textwidth]{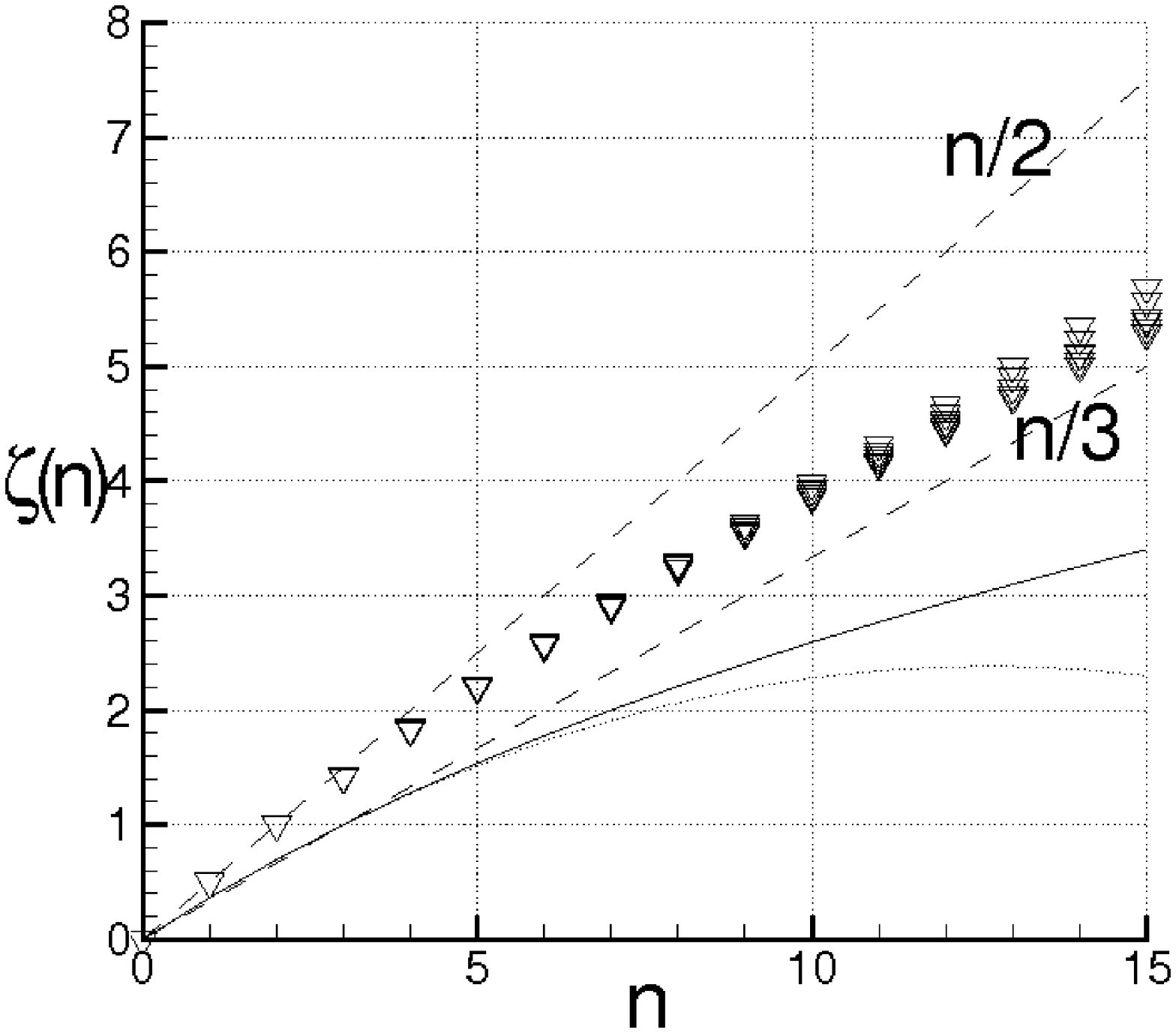}
\caption{Scaling exponents of $\Delta u_r$ for several $R_T$. Solid symbols are for the present data. Dashed line is for \cite{Kolmogorov41}. Dotted line is for \cite{Kolmogorov62}. 
 Continuous line is for \cite{She-Leveque94}}
\label{figura_9}
\end{figure}
%According to the present theory, $H_3(0) = -3/7 \approx -0.43$, against the 
%value of $-0.47$ calculated by \cite{Kim}. 
Although $H_3(0)$ does not depend upon the Reynolds number 
(\cite{deDivitiis_1}) (see Appendix), for $r > \lambda_T$, $H_3(\hat{r})$ rises with $R_T$ and goes to zero for $r \rightarrow \infty$. 
The constancy of $H_3(0)$ and the quadratic terms of Eq. (\ref{fluc4}) cause that the intermittency of $\Delta u_r$  increases with $R_T$, whereas
the spatial variations of $H_3$ and $H_4$ are the result of $f(r)$, $k(r)$ and
of  Eq. (\ref{fluc4}), thus also the scales of variation of skewness and flatness
are significantly smaller in comparison with those of the isotropic turbulence.
The quantity $H_4 - 3$ is significantly greater than zero for $r=0$ and tends to zero as 
$r \rightarrow \infty$ more rapidly than $H_3$.
\suppressfloats
\begin{figure}[t]
	\centering
         \includegraphics[width=0.65 \textwidth]{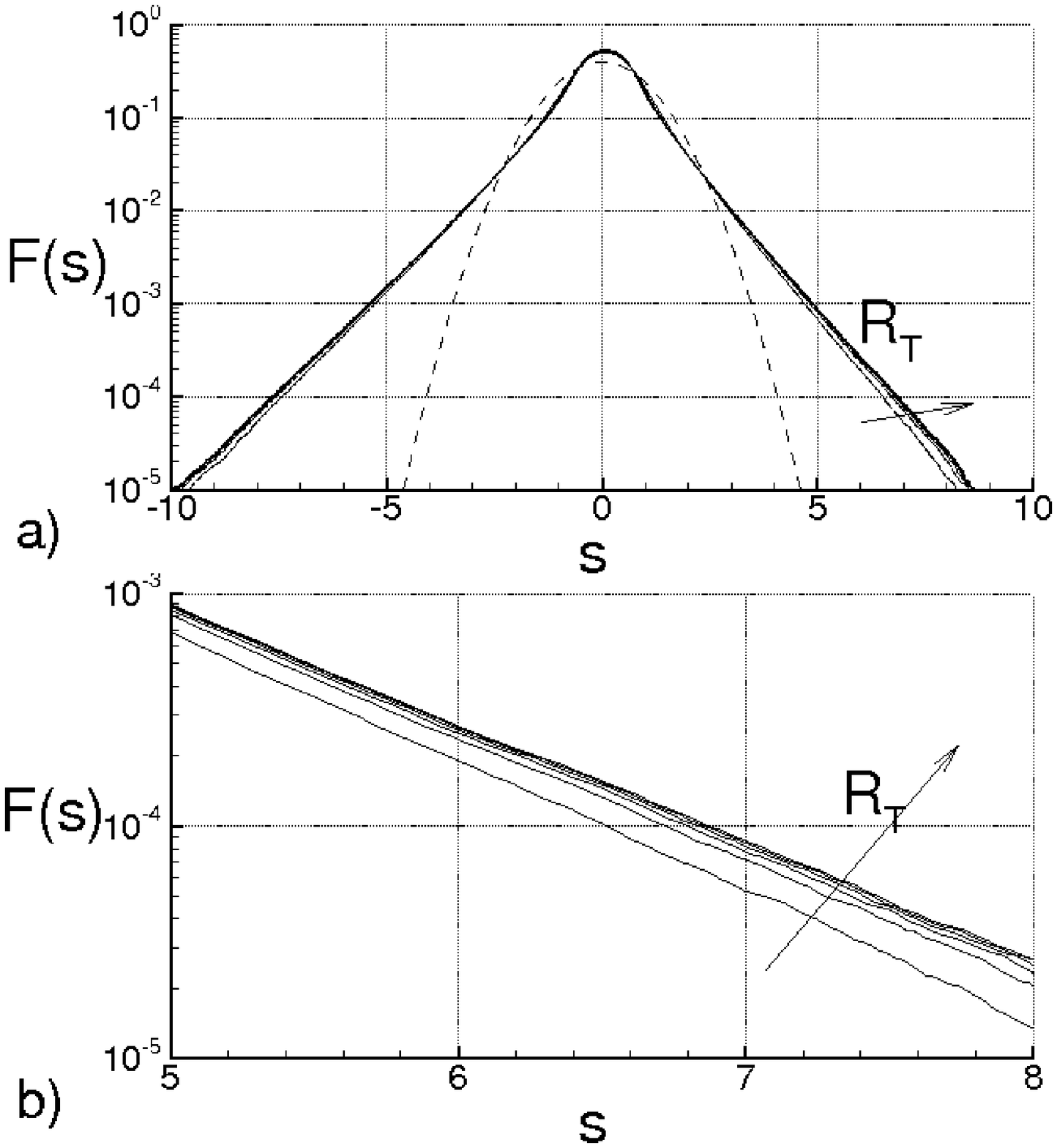}
\caption{Probability distribution functions of the longitudinal velocity derivative 
 for the different Taylor-Scale Reynolds numbers}
\label{figura_10}
\end{figure}

\bigskip

The spatial structure of $\Delta u_r$ is studied with the analysis 
reported in the Appendix (\cite{deDivitiis_1}), using the previous results.
%The analysis presented in Ref. \cite{deDivitiis_1} (see Appendix) 
%allows  to study the spatial structure of $\Delta u_r$, using the
% previous results. 
The statistical moments of $\Delta u_r(r)$ are expressed in function of $r$ through different scaling exponents $\zeta_n$
\bea
\left\langle (\Delta u_r)^{n} \right\rangle  = A_n r^{\zeta_n}
\label{fractal}
\eea
\suppressfloats
\begin{table}[t]
\centering
\caption{Scaling exponents of the longitudinal velocity difference for several Taylor-Scale Reynolds number.}
\begin{tabular}{ccccccc} 
$R_T$           &  100  &  200 &  300  &  400  & 500  &  600 \\
\hline
\hline \\
$\zeta_1$  &  0.49 & 0.50 &  0.50 &  0.50 & 0.50 &  0.50    \\ 
$\zeta_2$  &  1.00 & 1.00 &  1.00 &  1.00 & 1.00 &  1.00    \\ 
$\zeta_3$  &  1.42 & 1.41 &  1.41 &  1.41 & 1.41 &  1.41    \\
$\zeta_4$  &  1.84 & 1.83 &  1.82 &  1.82 & 1.83 &  1.83    \\
$\zeta_5$  &  2.22 & 2.19 &  2.19 &  2.19 & 2.20 &  2.20    \\
$\zeta_6$  &  2.59 & 2.56 &  2.56 &  2.56 & 2.57 &  2.58    \\
$\zeta_7$  &  2.94 & 2.89 &  2.90 &  2.90 & 2.91 &  2.93    \\
$\zeta_8$  &  3.28 & 3.23 &  3.23 &  3.24 & 3.26 &  3.29    \\
$\zeta_9$  &  3.60 & 3.54 &  3.55 &  3.56 & 3.59 &  3.63    \\
$\zeta_{10}$ &  3.92 & 3.85 &  3.87 &  3.88 & 3.93 &  3.98    \\
$\zeta_{11}$ &  4.22 & 4.15 &  4.17 &  4.19 & 4.26 &  4.32    \\
$\zeta_{12}$ &  4.53 & 4.44 &  4.47 &  4.51 & 4.58 &  4.66    \\
$\zeta_{13}$ &  4.81 & 4.72 &  4.75 &  4.81 & 4.91 &  5.00    \\
$\zeta_{14}$ &  5.10 & 5.00 &  5.04 &  5.11 & 5.24 &  5.35    \\
$\zeta_{15}$ &  5.39 & 5.28 &  5.33 &  5.42 & 5.56 &  5.69    \\
\hline
 \end{tabular}
\label{table2}
\end{table} 
In order to calculate $\zeta_n$,  $\langle \Delta u_r^n \rangle$ are first
calculated in function of $\hat{r}$, through Eqs. (\ref{m1}) 
(see for instance Fig.  \ref{figura_8}).
The scaling exponents are identified through a best fitting procedure, in the intervals ($\hat{r}_1, \hat{r}_2$), where the endpoints $\hat{r}_1$ and $\hat{r}_2$  have to be determined.
The calculation of $\zeta_n$ and $A_n$ is carried out through a minimum square method which, for each statistical moment, is applied to the following optimization problem
\bea
\ds J_n(\zeta_n, A_n) \hspace{-1.mm} \equiv  
\int_{\hat{r}_1}^{\hat{r}_2} 
\ds ( \langle (\Delta u_r)^n \rangle - A_n r^{\zeta_n} )^2 dr 
 = \mbox{min}, \   n = 1, 2, ...
\eea
where $\langle (\Delta u_r)^n \rangle$ are calculated with Eqs. (\ref{m1}),
$\hat{r}_1$ is assumed to be equal to $0.1$, whereas $\hat{r}_2$ is taken in such a way that $\zeta_2$ = 1 for all the values of $R_T$.
The so obtained scaling exponents, shown in Table \ref{table2}, are compared in Fig. \ref{figura_9} (solid symbols) with those of the Kolmogorov theories K41 (\cite{Kolmogorov41}) (dashed line) and K62 (\cite{Kolmogorov62}) (dotted line), and with the exponents calculated by  \cite{She-Leveque94} (continuous curve).
We found that, near the origin $\zeta_n \simeq n/2$ instead of the expected law $\zeta_n \simeq n/3$, whereas for $n>4$, $\zeta_n$ behaves like a multiscaling exponent.
This is in agreement with the experimental results of \cite{Swinney2010},
where the turbulence is produced in a cylindrical rotating tank.
The multiscaling behavior is the consequence of the combined effect of the quadratic terms of Eq. (\ref{fluc4}) and of the mechanism of the energy cascade expressed by $K$ through Eqs. (\ref{lyap}).

\bigskip

Next, to analyse the statistics of $\partial u_r /\partial {\hat{r}}$, 
the distribution functions of $\partial u_r /\partial {\hat{r}}$
are calculated using Eqs. (\ref{frobenious_perron}) and (\ref{fluc4}).
These are obtained with sequences of the variables $\xi$, $\eta$ and $\zeta$ generated by gaussian random numbers generators.
The distribution functions are then calculated through the statistical elaboration of these data and Eq. (\ref{fluc4}).
The results are shown in Fig. \ref{figura_10}a and \ref{figura_10}b in terms of the dimensionless abscissa 
\bea
\ds s = \frac{\partial u_r /\partial {\hat{r}} } 
{ \langle \left( \partial u_r /\partial {\hat{r}} \right) ^2 \rangle^{1/2}  }
\nonumber
\eea
where the dashed curve represents the gaussian PDF.
These distribution functions are normalized, in order that their standard 
deviations are equal to the unity.
In particular, Fig. \ref{figura_10}b shows the enlarged region of Fig. \ref{figura_10}a, where $5 < s < 8$. 
The tails of the PDFs change with $R_T$ in such a way that the intermittency rises with $R_T$ according to Eq. (\ref{fluc4}). 

\bigskip

\section{\bf  Conclusions}

The equation of the steady spherical correlation function is obtained in case of homogeneous turbulence in the presence of an uniform average velocity gradient, and an expression of the velocity correlation tensor in terms of this gradient is derived.
When $r=0$, this tensor recovers the Boussinesq closure of the Reynolds stress.
The solutions of this equation, calculated with the closure based on a specific Lyapunov theory previously proposed, allows to determine the statistics of $\Delta u_r$.
In particular:
\begin{itemize}
\item The energy spectrum, satisfying the continuity equation, follows the law 
$\kappa^{-2}$ in the inertial subrange whose size increases with the
   Reynolds number. This contrasts with the Kolmogorov law and represents a more
   tenuous mechanism of the energy cascade.
   Accordingly, the scaling exponents of the moments of velocity difference
 vary according to $\zeta_n \approx n/2$ instead of $\zeta_n \approx n/3$, and 
for $n>4$, these exponents exhibit multiscaling behavior.
\item The effect of the average velocity gradient, going against the mechanism
of energy cascade, makes the integral scales of $f$ and the scales of variations of $k$, significantly smaller than those calculated for the isotropic turbulence.
\item In case of uniform shear rate, the spanwise
correlation function of the streamwise velocity component,  exhibits the typical shape caused by 
the streaky coherent structures due to the vortex stretching. 
\end{itemize}
These results, which represent a further application of the analysis presented in  \cite{deDivitiis_1} and \cite{deDivitiis_2}, are comparable with the direct simulations of \cite{Kim} and are in agreement with the theoretical arguments of \cite{Gordienko01} and with the experiments presented in \cite{Swinney2010}.

\bigskip

\section{\bf Appendix: Statistics of velocity difference}

In this appendix, we recall the main results of \cite{deDivitiis_1}.
For sake of simplicity, we only consider the statistics of the longitudinal velocity difference $\Delta u_r$ associated to $R_{i k 0}$ (or $f$) and to Eq. (\ref{I2}).
This approximation allows to calculate all the dimensionless statistical
moments of $\Delta u_r$ once known the skewness $H_3(r)$.
This latter is calculated in terms of $f$ and $k$ (\cite{Batchelor53})
\bea
\ds H_3(r) = \frac{\left\langle ( \Delta u_r )^3 \right\rangle} 
{\left\langle (\Delta u_r)^2\right\rangle^{3/2}} =
  \frac{6 k(r)}{\left( 2 (1 -f(r)  )   \right)^{3/2} }
\label{H_3_01}
\eea
hence, $H_3(0) = -3/7$ does not depend upon the Reynolds number (\cite{deDivitiis_1}), and the higher order moments are consequentely determined, taking into account that 
$\Delta u_r$ is analytically expressed as  
\bea
\begin{array}{l@{\hspace{+0.2cm}}l}
\ds \frac {\Delta {u}_r}{\sqrt{\langle (\Delta {u}_r)^2} \rangle} =
\ds \frac{   {\xi} + \psi \left( \chi ( {\eta}^2-1 )  -  
\ds  ( {\zeta}^2-1 )  \right) }
{\sqrt{1+2  \psi^2 \left( 1+ \chi^2 \right)} } 
\end{array}
\label{fluc4}
\eea 
Equation (\ref{fluc4}), arising from statistical considerations about the 
Fourier-transformed Navier-Stokes equations, 
expresses the internal structure of the fully developed turbulence, 
where $\xi$, ${\eta}$ and $\zeta$ are independent centered random variables, each distributed following the gaussian distribution functions 
$p(\xi)$, $p(\eta)$ and $p(\zeta)$ whose standard deviation is equal to the unity.
The moments of $\Delta {u}_r$ are easily calculated from Eq. (\ref{fluc4}) 
\bea
\begin{array}{l@{\hspace{+0.2cm}}l}
\ds H_n \equiv \frac{\left\langle (\Delta u_r)^n \right\rangle}
{\left\langle (\Delta  u_r)^2\right\rangle^{n/2} }
= 
\ds \frac{1} {(1+2  \psi^2 \left( 1+ \chi^2 \right))^{n/2}} \\\\
\ds \sum_{k=0}^n 
\left(\begin{array}{c}
n  \\
k
\end{array}\right)  \psi^k
% \binom{n}{k} \psi^k
 \langle \xi^{n-k} \rangle 
  \langle (\chi(\eta^2 -1) - (\zeta^2 -1 ) )^k \rangle 
\end{array}
\label{m1}
\eea
where
\bea
\begin{array}{l@{\hspace{+0.2cm}}l}
\ds   \langle (\chi(\eta^2 -1) - (\zeta^2 -1 ) )^k \rangle = 
\ds \sum_{i=0}^k 
\left(\begin{array}{c}
k  \\
i
\end{array}\right)  
% \binom{k}{i} 
(-\chi)^i 
 \langle (\zeta^2 -1 )^i \rangle 
 \langle (\eta^2 -1 )^{k-i} \rangle, \\\\
\ds  \langle (\eta^2 -1 )^{i} \rangle = 
\sum_{l=0}^i 
%\binom{i}{l} 
\left(\begin{array}{c}
i  \\
l
\end{array}\right)  
(-1)^{l}
\langle \eta^{2(i-l)} \rangle 
 \end{array}
\label{m2}
\eea
In particular, $H_3$, related to the energy cascade, is
\bea
\ds H_3= \frac{  8  \psi^3 \left( \chi^3 - 1 \right) }
 {\left( 1+2  \psi^2 \left( 1+ \chi^2 \right) \right)^{3/2}  }
\label{H_3}
\eea 
and $\psi$ is a function of $r$ and of the Reynolds number \cite{deDivitiis_1}
\bea
\psi({\bf r}, R_T) =  
\sqrt{\frac{R_T}{15 \sqrt{15}}} \
\hat{\psi}(r)
\label{Rl}
\eea
being $\hat{\psi}(r)$ determined through Eq. (\ref{H_3}) as soon as $H_3(r)$ is known.
The parameter $\chi$ is also a function of $R_T$ and is implicitly calculated 
putting $r=0$ into Eq. (\ref{H_3}) 
\bea
\frac{  8  {\psi_0}^3 \left(  1-\chi^3 \right) }
 {\left( 1+2  {\psi_0}^2 \left( 1+ \chi^2 \right) \right)^{3/2}  }
= \frac{3}{7}
\label{sk1}
\eea
where ${\psi_0} = \psi(R_T,0)$ and $\hat{\psi_0} = 1.075$ \cite{deDivitiis_1}.
%\\
%From Eqs. (\ref{fluc4}) and (\ref{Rl}), all the quantities 
%$\vert  H_p(r) \vert$, for $p>3$,  rise with $R$, indicating that
% the intermittency of $\Delta u_r$ increases with the Reynolds
% number.

The PDF of $\Delta u_r$ can be formally expressed through the gaussian PDFs 
$p(\xi)$, $p(\eta)$ and $p(\zeta)$, using the Frobenius-Perron equation (\cite{Nicolis95})
\bea
\begin{array}{l@{\hspace{+0.3cm}}l}
F(\Delta {u'}_r) = 
\ds \int_\xi 
\int_\eta  
\int_\zeta 
p(\xi) p(\eta) p(\zeta) \
\delta \left( \Delta u_r-\Delta {u'}_r \right)   
\ d \xi d \eta d \zeta
\end{array}
\label{frobenious_perron}
\eea

\end{document}